\input phyzzx
\catcode`@=11 
\def\space@ver#1{\let\@sf=\empty \ifmmode #1\else \ifhmode
   \edef\@sf{\spacefactor=\the\spacefactor}\unskip${}#1$\relax\fi\fi}
\def\attach#1{\space@ver{\strut^{\mkern 2mu #1} }\@sf\ }
\newtoks\foottokens
\newbox\leftpage \newdimen\fullhsize \newdimen\hstitle
\newdimen\hsbody
\newif\ifreduce  \reducefalse
\def\almostshipout#1{\if L\lr \count2=1
      \global\setbox\leftpage=#1 \global\let\lr=R
  \else \count2=2
    \shipout\vbox{\special{dvitops: landscape}
      \hbox to\fullhsize{\box\leftpage\hfil#1}} \global\let\lr=L\fi}
\def\smallsize{\relax
\font\eightrm=cmr8 \font\eightbf=cmbx8 \font\eighti=cmmi8
\font\eightsy=cmsy8 \font\eightsl=cmsl8 \font\eightit=cmti8
\font\eightt=cmtt8
\def\eightpoint{\relax
\textfont0=\eightrm  \scriptfont0=\sixrm
\scriptscriptfont0=\sixrm
\def\rm{\fam0 \eightrm \f@ntkey=0}\relax
\textfont1=\eighti  \scriptfont1=\sixi
\scriptscriptfont1=\sixi
\def\oldstyle{\fam1 \eighti \f@ntkey=1}\relax
\textfont2=\eightsy  \scriptfont2=\sixsy
\scriptscriptfont2=\sixsy
\textfont3=\tenex  \scriptfont3=\tenex
\scriptscriptfont3=\tenex
\def\it{\fam\itfam \eightit \f@ntkey=4 }\textfont\itfam=\eightit
\def\sl{\fam\slfam \eightsl \f@ntkey=5 }\textfont\slfam=\eightsl
\def\bf{\fam\bffam \eightbf \f@ntkey=6 }\textfont\bffam=\eightbf
\scriptfont\bffam=\sixbf   \scriptscriptfont\bffam=\sixbf
\def\tt{\fam\ttfam \eightt \f@ntkey=7 }
\def\caps{\fam\cpfam \tencp \f@ntkey=8 }\textfont\cpfam=\tencp
\setbox\strutbox=\hbox{\vrule height 7.35pt depth 3.02pt width\z@}
\samef@nt}
\def\Eightpoint{\eightpoint \relax
  \ifsingl@\subspaces@t2:5;\else\subspaces@t3:5;\fi
  \ifdoubl@ \multiply\baselineskip by 5
            \divide\baselineskip by 4\fi }
\parindent=16.67pt
\itemsize=25pt
\thinmuskip=2.5mu
\medmuskip=3.33mu plus 1.67mu minus 3.33mu
\thickmuskip=4.17mu plus 4.17mu
\def\thinspace{\kern .13889em }
\def\negthinspace{\kern-.13889em }
\def\enspace{\kern.416667em }
\def\enskip{\hskip.416667em\relax}
\def\quad{\hskip.83333em\relax}
\def\qquad{\hskip1.66667em\relax}
\def\crr{\cropen{8.3333pt}}
\foottokens={\Eightpoint\singlespace}
\def\papersize{\SIZE\OFFSET\skip\footins=\bigskipamount}
\def\SIZE{\hsize=11.8truecm\vsize=17.5truecm}
\def\OFFSET{\voffset=-1.3truecm\hoffset=  .14truecm}
\message{STANDARD CERN-PREPRINT FORMAT}
\def\attach##1{\space@ver{\strut^{\mkern 1.6667mu ##1} }\@sf\ }
\def\PH@SR@V{\doubl@true\baselineskip=20.08pt plus .1667pt minus
.0833pt
             \parskip = 2.5pt plus 1.6667pt minus .8333pt }
\def\author##1{\vskip\frontpageskip\titlestyle{\tencp ##1}\nobreak}
\def\address##1{\par\kern 4.16667pt\titlestyle{\tenpoint\it ##1}}
\def\andaddress{\par\kern 4.16667pt \centerline{\sl and} \address}
\def\abstract{\vskip2\frontpageskip\centerline{\tenrm Abstract}
              \vskip\headskip }
\def\cases##1{\left\{\,\vcenter{\Tenpoint\m@th
    \ialign{$####\hfil$&\quad####\hfil\crcr##1\crcr}}\right.}
\def\matrix##1{\,\vcenter{\Tenpoint\m@th
    \ialign{\hfil$####$\hfil&&\quad\hfil$####$\hfil\crcr
      \mathstrut\crcr\noalign{\kern-\baselineskip}
     ##1\crcr\mathstrut\crcr\noalign{\kern-\baselineskip}}}\,}
\Tenpoint
}
\def\Smallsize{\smallsize\reducetrue
\let\lr=L
\hstitle=8truein\hsbody=4.75truein\fullhsize=24.6truecm\hsize=\hsbody
\output={
  \almostshipout{\leftline{\vbox{\makeheadline
  \pagebody\makefootline}}}\advancepageno
     }
\special{dvitops: landscape}
\def\makeheadline{
\iffrontpage\line{\the\headline}
             \else\vskip .0truecm\line{\the\headline}\vskip .5truecm
\fi}
\def\makefootline{\iffrontpage\vskip  0.truecm\line{\the\footline}
               \vskip -.15truecm\line{\the\date\hfil}
              \else\line{\the\footline}\fi}
\paperheadline={
\iffrontpage\hfil
               \else
               \tenrm\hss $-$\ \folio\ $-$\hss\fi    }
\paperstyle}
%
%
%
%
%
%
%
%
%
\newcount\referencecount     \referencecount=0
\newif\ifreferenceopen       \newwrite\referencewrite
\newtoks\rw@toks
\def\NPrefmark#1{\attach{\scriptscriptstyle [ #1 ] }}
\let\PRrefmark=\attach
\def\refmark#1{\relax\ifPhysRev\PRrefmark{#1}\else\NPrefmark{#1}\fi}
\def\refend{\refmark{\number\referencecount}}
\newcount\lastrefsbegincount \lastrefsbegincount=0
\def\refsend{\refmark{\count255=\referencecount
   \advance\count255 by-\lastrefsbegincount
   \ifcase\count255 \number\referencecount
   \or \number\lastrefsbegincount,\number\referencecount
   \else \number\lastrefsbegincount-\number\referencecount \fi}}
\def\refch@ck{\chardef\rw@write=\referencewrite
   \ifreferenceopen \else \referenceopentrue
   \immediate\openout\referencewrite=referenc.texauxil \fi}
%
{\catcode`\^^M=\active 
  \gdef\obeyendofline{\catcode`\^^M\active \let^^M\ }}%
%
{\catcode`\^^M=\active 
  \gdef\ignoreendofline{\catcode`\^^M=5}}
{\obeyendofline\gdef\rw@start#1{\def\t@st{#1} \ifx\t@st\blankend%
\endgroup \@sf \relax \else \ifx\t@st\bl@nkend \endgroup \@sf \relax%
\else \rw@begin#1
\backtotext
\fi \fi } }
{\obeyendofline\gdef\rw@begin#1
{\def\n@xt{#1}\rw@toks={#1}\relax%
\rw@next}}
\def\blankend{}
{\obeylines\gdef\bl@nkend{
}}
\newif\iffirstrefline  \firstreflinetrue
\def\rwr@teswitch{\ifx\n@xt\blankend \let\n@xt=\rw@begin %
 \else\iffirstrefline \global\firstreflinefalse%
\immediate\write\rw@write{\noexpand\obeyendofline \the\rw@toks}%
\let\n@xt=\rw@begin%
      \else\ifx\n@xt\rw@@d \def\n@xt{\immediate\write\rw@write{%
        \noexpand\ignoreendofline}\endgroup \@sf}%
             \else \immediate\write\rw@write{\the\rw@toks}%
             \let\n@xt=\rw@begin\fi\fi \fi}
\def\rw@next{\rwr@teswitch\n@xt}
\def\rw@@d{\backtotext} \let\rw@end=\relax
\let\backtotext=\relax

\newdimen\refindent     \refindent=30pt
\def\refitem#1{\par \hangafter=0 \hangindent=\refindent
\Textindent{#1}}
\def\REFNUM#1{\space@ver{}\refch@ck \firstreflinetrue%
 \global\advance\referencecount by 1 \xdef#1{\the\referencecount}}
\def\refnum#1{\space@ver{}\refch@ck \firstreflinetrue%
 \global\advance\referencecount by 1
\xdef#1{\the\referencecount}\refend}

\def\REF#1{\REFNUM#1%
 \immediate\write\referencewrite{%
 \noexpand\refitem{#1.}}%
\begingroup\obeyendofline\rw@start}
\def\ref{\refnum\?%
 \immediate\write\referencewrite{\noexpand\refitem{\?.}}%
\begingroup\obeyendofline\rw@start}
\def\Ref#1{\refnum#1%
 \immediate\write\referencewrite{\noexpand\refitem{#1.}}%
\begingroup\obeyendofline\rw@start}
\def\REFS#1{\REFNUM#1\global\lastrefsbegincount=\referencecount
\immediate\write\referencewrite{\noexpand\refitem{#1.}}%
\begingroup\obeyendofline\rw@start}
\newif\ifmref  
\newif\iffref  
\def\xrefsend{\xrefmark{\count255=\referencecount
\advance\count255 by-\lastrefsbegincount
\ifcase\count255 \number\referencecount
\or \number\lastrefsbegincount,\number\referencecount
\else \number\lastrefsbegincount-\number\referencecount \fi}}
\def\xrefsdub{\xrefmark{\count255=\referencecount
\advance\count255 by-\lastrefsbegincount
\ifcase\count255 \number\referencecount
\or \number\lastrefsbegincount,\number\referencecount
\else \number\lastrefsbegincount,\number\referencecount \fi}}
\def\xREFNUM#1{\space@ver{}\refch@ck\firstreflinetrue%
\global\advance\referencecount by 1
\xdef#1{\xrefend}}
\def\xrefend{\xrefmark{\number\referencecount}}
\def\xrefmark#1{[{#1}]}
\def\xRef#1{\xREFNUM#1\immediate\write\referencewrite%
{\noexpand\refitem{#1 }}\begingroup\obeyendofline\rw@start}%
\def\xREFS#1{\xREFNUM#1\global\lastrefsbegincount=\referencecount%
\immediate\write\referencewrite{\noexpand\refitem{#1 }}%
\begingroup\obeyendofline\rw@start}
\def\rrr#1#2{\relax\ifmref{\iffref\xREFS#1{#2}%
\else\xRef#1{#2}\fi}\else\xRef#1{#2}\xrefend\fi}
\def\doubref#1#2{\mreftrue\freftrue{#1}%
\freffalse{#2}\mreffalse\xrefsdub}
\referencecount=0
\def\refbreak{\hfil\penalty200\hfilneg}
\def\paperstyle{\papers}
\paperstyle   
%
%
%
\def\slacpub{\afterassignment\slacp@b\toks@}
\def\slacp@b{\edef\n@xt{\Pubnum={\the\toks@}}\n@xt}
\let\pubnum=\slacpub
\expandafter\ifx\csname eightrm\endcsname\relax
    \let\eightrm=\ninerm \let\eightbf=\ninebf \fi

\font\seventeencp=cmcsc10 scaled\magstep3

\newif\ifCONF \CONFfalse
\newif\ifBREAK \BREAKfalse
\newif\ifsectionskip \sectionskiptrue

%
%
%
%
\def\NuclPhysProc{
\let\lr=L
\hstitle=8truein\hsbody=4.75truein\fullhsize=21.5truecm\hsize=\hsbody
\hstitle=8truein\hsbody=4.75truein\fullhsize=20.7truecm\hsize=\hsbody
\output={
  \almostshipout{\leftline{\vbox{\makeheadline
  \pagebody\makefootline}}}\advancepageno
     }
\def\papersize{\SIZE\OFFSET\skip\footins=\bigskipamount}
\def\SIZE{\hsize=10.0truecm\vsize=27.0truecm}
\def\OFFSET{\voffset=-1.4truecm\hoffset=-2.40truecm}
\message{NUCLEAR PHYSICS PROCEEDINGS FORMAT}
\def\makeheadline{
\iffrontpage\line{\the\headline}
             \else\vskip .0truecm\line{\the\headline}\vskip .5truecm
\fi}
\def\makefootline{\iffrontpage\vskip  0.truecm\line{\the\footline}
               \vskip -.15truecm\line{\the\date\hfil}
              \else\line{\the\footline}\fi}
\paperheadline={\hfil}
\paperstyle}
%
%
%
%

%
%
%
%

%
%
%
%
\def\ReprintVolume{\smallsize
\def\papersize{\hsize=18.0truecm\vsize=23.1truecm\voffset -.73truecm
    \hoffset -.65truecm\skip\footins=\bigskipamount
    \normaldisplayskip= 20pt plus 5pt minus 10pt}
\message{REPRINT VOLUME FORMAT}
\paperstyle\baselineskip=.425truecm\parskip=0truecm
\def\makeheadline{
\iffrontpage\line{\the\headline}
             \else\vskip .0truecm\line{\the\headline}\vskip .5truecm
\fi}
\def\makefootline{\iffrontpage\vskip  0.truecm\line{\the\footline}
               \vskip -.15truecm\line{\the\date\hfil}
              \else\line{\the\footline}\fi}
\paperheadline={
\iffrontpage\hfil
               \else
               \tenrm\hss $-$\ \folio\ $-$\hss\fi    }
\def\sectionfont{\bf}    }
%
%
%
%
\def\SIZE{\hsize=15.73truecm\vsize=23.11truecm}
\def\OFFSET{\voffset=0.0truecm\hoffset=0.truecm}
\message{DEFAULT FORMAT}
\def\papersize{\SIZE\OFFSET\skip\footins=\bigskipamount
\normaldisplayskip= 35pt plus 3pt minus 7pt}
\Pubnum={\rm \the\pubnum }
\def\title#1{\vskip\frontpageskip\vskip .50truein
     \titlestyle{\seventeencp #1} \vskip\headskip\vskip\frontpageskip
     \vskip .2truein}
\def\author#1{\vskip .27truein\titlestyle{#1}\nobreak}

\def\p@bblock{\begingroup \tabskip=\hsize minus \hsize
   \baselineskip=1.5\ht\strutbox \topspace+2\baselineskip
   \halign to\hsize{\strut ##\hfil\tabskip=0pt\crcr
  \the \Pubnum\cr}\endgroup}
\def\makefootline{\iffrontpage\vskip .27truein\line{\the\footline}
                 \vskip -.1truein
              \else\line{\the\footline}\fi}
\paperfootline={\iffrontpage\message{FOOTLINE}
\hfil\else\hfil\fi}

\def\abstract{\vskip2\frontpageskip\centerline{\twelvebf Abstract}
              \vskip\headskip }

\paperheadline={
\iffrontpage\hfil
               \else
               \twelverm\hss $-$\ \folio\ $-$\hss\fi}
%
%
\def\nup#1({\refbreak\ Nucl.\ Phys.\ $\underline {B#1}$\ (}
\def\plt#1({\refbreak\ Phys.\ Lett.\ $\underline  {#1}$\ (}
\def\cmp#1({\refbreak\ Commun.\ Math.\ Phys.\ $\underline  {#1}$\ (}
\def\prp#1({\refbreak\ Physics\ Reports\ $\underline  {#1}$\ (}
\def\prl#1({\refbreak\ Phys.\ Rev.\ Lett.\ $\underline  {#1}$\ (}
\def\prv#1({\refbreak\ Phys.\ Rev. $\underline  {D#1}$\ (}
\def\und#1({            $\underline  {#1}$\ (}
%
%

\def\rB{\hfil\penalty1000\hfilneg}
%
%
\hyphenation{sym-met-ric anti-sym-me-tric re-pa-ra-me-tri-za-tion
Lo-rentz-ian a-no-ma-ly di-men-sio-nal two-di-men-sio-nal}
%
%
%
%

\def\coeff#1#2{{\textstyle { #1 \over #2}}\displaystyle}
\def\boxit#1{\vbox{\hrule\hbox{\vrule\kern3pt
\vbox{\kern3pt#1\kern3pt}\kern3pt\vrule}\hrule}}
\message{ by V.K, W.L and A.S}
\catcode`@=12
\paperstyle
\input tables

\def\ai  {(a,i)}
\def\bi  {(b,i)}
\def\bj  {(b,j)}
\def\chi {\X}
\def\ck  {(c,k)}
\def\complex{{\bf C}}
\def\ee  {{\rm e}}
\def\el  {\ell}
\def\GG  {{\,;\G}}
\def\Gt  {{\tilde A_\G}}
\def\HH  {{\,;\H}}
\def\HHJ {{\,;\H_J}}
\def\id  {{\sl id}}
\def\ii  {{\rm i}}
\def\ip  {{i_\el}}
\def\one {{\bf 1}}
\def\P   {{\cal P}}
\def\Pgg {{\cal P}_{\Gamma,\Gamma}}
\def\Pgs {{\cal P}_{\Gamma,S}}
\def\Psg {{\cal P}_{S,\Gamma}}
\def\Pss {{\cal P}_{S,S}}
\def\SO  {{\rm so}}
\def\SU  {{\rm su}}
\def\sumRQ#1{\sum_{\scriptstyle #1 \atop Q=0;\,{\rm R}}}
\def\Tau {{\cal T}}
\def\tims{\times}
\def\unit{{\hbox{{\tenpoint 1}$\!\!$1}}}
\def\Xt  {\tilde{\cal X}}
\def\FormOne  {(2.5)}
\def\XX       {(5.1)}
\def\PhaseRot {(5.2)}
\def\SJFSJ    {(5.6)}
\def\FonOrbit {(A.1)}
\def\Fformh   {(A.13)}
\def\Fbar     {(A.14)}
\def\ScYA{\rrr\ScYA{
A.N.~Schellekens and S.~Yankielowicz,
\nup 327 (1989) 673; \rB
\plt B227 (1989) 387.}}
\def\ScYe{\rrr\ScYe{A.N.~Schellekens and S.~Yankielowicz,
\nup334 (1990) 67.}}
\def\ScYg{\rrr\ScYg{A.N.~Schellekens and S.~Yankielowicz,
Int.~J.~Mod.~Phys.\und{A5} (1990) 2903.}}
\def\FSch{\rrr\FSch {J.~Fuchs and C.~Schweigert,
Ann.~Phys.\und{234}(1994) 102.}}
\def\FSS{\rrr\FSS{ J.~Fuchs, A.N.~Schellekens and C.~Schweigert,
{\it Galois Modular Invariants of WZW-models},
\nup{437} (1995) 667.}}
\def\FSSG{\rrr\FSSG{ J.~Fuchs, A.N.~Schellekens and C.~Schweigert,
{\it Quasi-Galois Symmetries of the Modular S-Matrix}
preprint NIKHEF-94/37; Commun.~Math.~Phys., in press.}}
\def\Alig{\rrr\Alig{K.~Intriligator,
\nup 332 (1990) 541.}}
\def\Mooresei{\rrr\Mooresei{G.\ Moore and N.\ Seiberg,
\cmp123 (1989) 177.}}
\def\MoSc{\rrr\MoSc{
G.~Moore   and N.~Seiberg, 
\plt B220 (1989) 422.}}
\def\BeBT{\rrr\BeBT {B.~Gato-Rivera and A.N.~Schellekens,
\cmp{145} (1992) 85.}}
\def\SchK{\rrr\SchK{M.~Kreuzer and A.N.~Schellekens,
\nup411 (1994) 97.}}
\def\EVeA{\rrr\EVeA{E.~Verlinde,
\nup300 (1988) 360.}}
\def\FSSc{\rrr\FSSc{ J.~Fuchs, A.N.~Schellekens and C.~Schweigert,
{\it From Dynkin Diagram Automorphisms to Fixed Point Structures},
preprint hep-th/9506135;\rB ~Commun.~Math.~Phys., in press.}}
\def\FSSb{\rrr\FSSb{J.~Fuchs, A.N.~Schellekens and C.~Schweigert,
{\it The resolution of field identification
           fixed points in diagonal coset theories}, preprint
           hep-th/9509105; Nucl.~Phys.~$B$, in press.}}
\def\FSSd{\rrr\FSSd{J.~Fuchs, A.N.~Schellekens and C.~Schweigert,
{\it Twining characters, orbit Lie algebras, and fixed point
resolution},
preprint q-alg/9511026; to appear in the Proceedings of the
{\it Workshop on New Trends in Quantum Field Theory}
(Razlog, Bulgaria, August 1995).}}
\def\DuJo{\rrr\DuJo{D.~Dunbar and K.~Joshi, Mod.~Phys.~Lett.\und{A8}
(1993)
 2803}}
\def\FBk{\rrr\FBk{J.~Fuchs, {\it Affine Lie Algebras and Quantum
Groups},
Cambridge University Press (1992).}}
\def\KaPe{\rrr\KaPe{
V.G.~Kac and D.H.~Peterson, Adv.~in Math.~53 (1984) 125.}}
\def\DVVV{\rrr\DVVV{R.~Dijkgraaf, C.~Vafa, E.~Verlinde and
H.~Verlinde,
\cmp 123 (1989) 16.}}
\def\fegk{\rrr\fegk{G.\ Felder, K.\ Gawedzki, and A.\ Kupiainen,
\cmp{117} (1988) 127.}}
%
%
\def\AbGe{\rrr\AbGe{A.\ Abouelsaood and D.\ Gepner,
\plt{176} (1986) 380.}}
\def\Kac{\rrr\Kac{
V.G.~Kac, {\it  Infinite Dimensional Lie Algebras}, third edition,
Cambridge University Press (1990).}}

\paperstyle

\def\GCD{{\rm gcd}}
\def\mod{{\rm ~mod~}}

\def\half{\coeff12}
\def\TwineChap{{6}}
\def\U{{\cal U}}
\def\M{{\cal M}}
\def\H{{\cal H}}
\def\Zbf{{\bf Z}}

\def\X{{\cal X}}
\def\G{{\cal G}}
\def\S{{\cal S}}

\catcode`@=11
\def\ninef@nts{\relax
    \textfont0=\ninerm          \scriptfont0=\sixrm
      \scriptscriptfont0=\sixrm
    \textfont1=\ninei           \scriptfont1=\sixi
      \scriptscriptfont1=\sixi
    \textfont2=\ninesy          \scriptfont2=\sixsy
      \scriptscriptfont2=\sixsy
    \textfont3=\tenex          \scriptfont3=\tenex
      \scriptscriptfont3=\tenex
    \textfont\itfam=\nineit     \scriptfont\itfam=\seveni  
\sevenit
    \textfont\slfam=\ninesl     \scriptfont\slfam=\sixrm 
\sevensl
    \textfont\bffam=\ninebf     \scriptfont\bffam=\sixbf
      \scriptscriptfont\bffam=\sixbf
    \textfont\ttfam=\tentt
    \textfont\cpfam=\tencp }
\def\ninepoint{\ninef@nts \samef@nt \b@gheight=9pt \setstr@t }
\newif\ifnin@  \nin@false
\def\Tenpoint{\tenpoint\twelv@false\nin@false\spaces@t}
\def\Twelvepoint{\twelvepoint\twelv@true\nin@false\spaces@t}
\def\Ninepoint{\ninepoint\twelv@false\nin@true\spaces@t}
\def\spaces@t{\rel@x
      \iftwelv@ \ifsingl@\subspaces@t3:4;\else\subspaces@t1:1;\fi
       \else \ifsingl@\subspaces@t3:5;\else\subspaces@t4:5;\fi \fi
      \ifdoubl@ \multiply\baselineskip by 5
         \divide\baselineskip by 4 \fi
       \ifnin@ \ifsingl@\subspaces@t3:8;\else\subspaces@t4:7;\fi \fi
}
\def\Vfootnote#1{\insert\footins\bgroup
   \interlinepenalty=\interfootnotelinepenalty \floatingpenalty=20000
   \singl@true\doubl@false \iftwelv@ \Tenpoint
   \else \Ninepoint \fi
   \splittopskip=\ht\strutbox \boxmaxdepth=\dp\strutbox
   \leftskip=\footindent \rightskip=\z@skip
   \parindent=0.5\footindent \parfillskip=0pt plus 1fil
   \spaceskip=\z@skip \xspaceskip=\z@skip \footnotespecial
   \Textindent{#1}\footstrut\futurelet\next\fo@t}

\def\small#1{\vskip .3truecm\footnoterule\nobreak
\Ninepoint\parindent=2pc\sl
\hang #1 \vskip .3truecm\nobreak\footnoterule}

\def\small#1{}

  

\pubnum={{}}
\rightline{DESY 96--008}
\rightline{IHES/P/96/8}
\rightline{NIKHEF/96-001}
\rightline{hep-th/9601078}
\rightline{January 1996}
\date{January 1996}
\pubtype{CRAP}
\titlepage
\message{TITLE}

\title{\fourteenbf A matrix $S$ for all simple current extensions}
\author{~~J. Fuchs\rlap,\foot{DESY, Notkestra\ss e 85,
D$\,-\,$22603 Hamburg, Germany}
A. N. Schellekens\rlap,\foot{NIKHEF, Postbus 41882, NL$\,-\,$1009
DB Amsterdam, The Netherlands.\hfill\break E-mail: t58@attila.nikhef.nl}
C. Schweigert\foot{IHES, 35 route de Chartres, F$\,-\,$91440
Bures-sur-Yvette, France}}
\bigskip

\abstract \noindent
A formula is presented for the modular transformation matrix
$S$ for any simple current extension of the chiral algebra
of a conformal field theory. This
   provides in particular an algorithm
for resolving arbitrary simple current fixed points,
in such a way that the matrix $S$
we obtain is unitary and symmetric and furnishes a modular group
representation. The formalism works in principle for any conformal
field theory. A crucial ingredient is a set of matrices $S^J_{ab}$,
where $J$ is a simple current and $a$ and $b$ are fixed points of $J$.
We expect that these input matrices realize the modular group for
the torus one-point functions of the simple currents.
In the case of WZW-models these matrices can be identified
with the $S$-matrices of the orbit Lie algebras that were
introduced recently in \FSSc.
As a special case of our conjecture we obtain the modular matrix
$S$ for WZW-theories based on group manifolds that are not simply
connected, as well as for most coset models.

   \baselineskip= 15.0pt plus .2pt minus .1pt

\chapter{Introduction}

One of the more important unsolved problems in conformal field
theory is that of classifying and understanding all modular invariant
partition functions. Besides the diagonal modular invariant,
one can often construct other modular invariant partition
functions for conformal field theories.
In spite of some recent progress, even in the most extensively
studied case of WZW-models based on simple Lie algebras, the
classification of these invariants is still incomplete.
Moreover, even for the known non-diagonal invariants, a satisfactory
interpretation
     as a full-fledged conformal field theory
is available in only a few cases.

     In this paper,
we will be interested in modular invariants that suggest an
extension of the chiral algebra, \ie\ invariants of the
general form\foot{In particular we do not study `heterotic'
invariants
or fusion rule automorphisms, since our interest is in defining
the matrix $S$ for the chiral half of a theory. }
  $$ \sum_i N_i\, | \sum_{\ell} m_{i,\ell} \chi_{\ell} |^2 \ .
  \eqn\Pfunct$$
Here $\chi_{\ell}$ is a character of the original theory
(which we will call the unextended theory, even though its chiral
algebra will in general itself be an extension of the Virasoro
algebra), $m_{i,\ell}$
a non-negative integer and $N_i$ a positive integer. The identity
character of the unextended theory appears exactly once (by
convention for $i=\ell=0$, with $m_{0,0}=N_0 =1$).

Such a partition function suggests an interpretation in terms of an
extended algebra, with each term representing the contribution
of an irreducible representation of that algebra.
The fields which we would like to interpret as
the generators of an extended chiral algebra
can then be read off the term containing the identity.
The existence and uniqueness of such an extended algebra is however by no
means guaranteed.
Indeed, several examples are known of partition
functions of the form \Pfunct\ that do not correspond to any
conformal field theory (see \eg\ \doubref\ScYe{\FSS}).
We are not aware of examples of modular invariant combinations
of characters of rational conformal field theories that can be
interpreted in more than one way in terms of an extended chiral
algebra, but this possibility cannot be ruled out either.

Having found a modular invariant partition function, the next logical
step is to attempt to derive the modular transformation
matrix $S$ of the characters of the putative new theory. If such a
matrix can indeed be written down, a further important consistency
check is the computation of the fusion coefficients using Verlinde's
formula \EVeA. If no inconsistency appears, one can try to compute
operator product coefficients and correlation functions. In
principle, any of these steps may fail or produce a non-unique answer.

Apart from a few trivial theories, essentially the only case
where the whole programme can be carried through is the extension of
WZW-models by currents of spin 1. These invariants can be interpreted
as conformal embeddings, and hence the extended theory is again a
WZW-model.

In this paper we will focus on another case that can be expected to
be manageable, namely, for arbitrary rational conformal field
theories, the so-called
{\it simple current invariants} ($\!\!$\doubref\ScYA{\Alig}, for a
review see \ScYg). These invariants have been completely
classified for any conformal
field theory \doubref\BeBT{\SchK}. Since the construction of the
partition function can be formulated in terms of orbifold
methods, it is reasonable to expect a conformal field theory
to exist. Therefore in particular there should exist
a unitary and symmetric matrix $S$ with all the usual properties.
Unfortunately, orbifold methods do not seem to be of much help
in actually determining this matrix. Such a computation has
been carried out so far only for the $\Zbf_2$-orbifolds
of the $c=1$ models \DVVV\ and a few other simple examples.
Therefore we will follow a different route. Here we will only
consider the first step in the programme of describing the (putative)
theory which corresponds to a given simple current modular
invariant, namely the determination of $S$.

Our current knowledge indicates that for WZW-models based on simple
Lie algebras nearly all off-diagonal invariants are simple current
invariants. The remaining solutions, which are appropriately referred
to as `exceptional invariants', are rare (although there are
a few infinite series) and unfortunately beyond the scope of this
paper. For semi-simple algebras far less is known, but certainly the
number of simple current invariants increases dramatically \BeBT.
For most of these invariants the modular matrix $S$, one
of the most basic quantities of a conformal field theory, could
not be computed up to now. Although the most important application
of our results appears to be in WZW-models, and also
in coset theories (see below),
we emphasize that simple current constructions are not {\it a priori}
restricted to WZW-models. For this reason we will set up the
formalism in its most general form, and focus on WZW-models only at
the end.

For simple current invariants there are a few convenient
simplifications in \Pfunct; for example the coefficients
$m_{i,\ell}$ are either 0 or 1, and
the vectors $\vec m_i$ are all orthogonal.
The problem we  address in this paper occurs whenever one of the
multiplicities $N_i$ is larger than 1. This situation occurs if
one (or more) of the simple currents in the extension has
a fixed point, \ie\ if it maps a primary field to itself.
If there are no fixed points, one can compute the matrix $S$ simply
by looking at the modular transformation properties of the characters.
However, if $N_i > 1$ for some value of $i$,
this may imply that the new theory has more than one character
corresponding to the $i^{\rm th}$ term (the multiplicity will
in fact be determined in this paper).
In that case
all characters in the $i$th term of the sum \Pfunct\
are identical as functions
of the modular parameter $\tau$ and possible Cartan angles of the
unextended  theory, and one cannot disentangle their transformation
under $\tau\mapsto -{1\over\tau}$.

Fixed points occur very often in simple current invariants. A simple
and well-known example is the $D$-invariant of $\SU(2)$ level 4,
which has the form $|\chi_0+\chi_4|^2 + 2\,|\chi_2 |^2$. There are two
representations with character $\chi_2$.  The known modular
transformations of $\SU(2)$ level 4 do not tell us how they transform
into each other.  Hence we cannot deduce the matrix $S$ directly from
that of $\SU(2)$ level 4. If we assume that a new theory with an extended
chiral algebra exists, we know more about $S$: it must be unitary and
symmetric and
form, together with the known matrix $T$, a representation of
the modular group, hence satisfy $S^4=\unit$ and $(ST)^3=S^2$. In the
example the most general form of $S$ that is symmetric and
agrees with the known transformations of the $\SU(2)_4$ characters is
  $${1\over \sqrt3}\pmatrix{
  1 & 1 &1 \cr  1 &-\half + \epsilon & -\half - \epsilon\cr
  1 & -\half - \epsilon & -\half + \epsilon  \cr }\ , $$
where $\epsilon$ is an unknown parameter. Imposing unitarity
fixes $\epsilon$ up to a sign. Finally imposing $(ST)^3=S^2$ fixes
$\epsilon$ completely (and one obtains the matrix $S$
of $\SU(3)$ level 1).  It is this solution that we wish to generalize
to arbitrary conformal field theories with simple currents.

In the general case one can proceed as follows \ScYe. First one
computes the naive matrix $S$ associated with
the partition function \Pfunct\ by `orbit-averaging'
the  matrix $S$ of the original theory,  and by resolving
the $i$th row and column of $S$ into (at most)
$N_i$ distinct rows and columns. To make the new matrix
unitary, correction terms are needed for the entries between
fixed point representations. These corrections can be described in
terms of a  matrix $S^J$ that acts only on the fixed
points (in fact there is such a matrix for every current $J$
in the extension, hence the upper index). It can then be shown that
the resolved matrix\foot{To prevent confusion between the matrices for
the unextended and the extended theories, we denote the former as $S$
and the latter as $\tilde S$.}
 $\tilde S$ is unitary and symmetric and
satisfies $(\tilde ST)^3=\tilde S^2$ if $S^J$  has all those
properties on the fixed points. Since
$T$ is known and unambiguous, this information can be used in some
cases to get plausible {\it ans\"atze} for $S^J$.

The problem with this method is that one has to identify
the $T$-eigenvalues of  the
degenerate representations with a known spectrum. Surprisingly,
in many WZW-models these $T$-eigenvalues can be  recognized as
those of another WZW-theory (up to an overall phase). In \ScYg\ this
was achieved for all simple current invariants of WZW-models based
on simple, simply laced Lie algebras, as well as for
a few other cases. However, the fixed point spectrum obtained for
$B_n$ and $C_{2n}$ theories did not correspond (with a few exceptions)
to that of a WZW-model or any other known conformal field theory.
In addition, the application of this procedure to more
complicated combinations of simple currents, with fixed
points of all possible types, has never been formulated.

\noindent
The main results reported here are:
\item\bullet
A conjecture is presented for the matrix $\tilde S$ for any simple
current invariant of any conformal field theory for which the
relevant matrices $S^J$ are known. One important problem to be
addressed is precisely how many irreducible
representations of the extended algebra one gets if $N_i > 1$.
We will present a conjecture for this case as well;
perhaps surprisingly, the answer is not always $N_i$. This means
in particular that not even the spectrum of certain extended
theories was known before.
\item\bullet
A matrix $S^J$ is presented for any simple current of any WZW-model.
This requires the extension of the results of \ScYg\ to all simple
algebras. The construction of $S^J$ was essentially already achieved in
\FSSc. It was found that the `missing' cases
correspond to  spectra of twisted affine Kac-Moody
algebras. The matrices $S^J$ for
the missing cases have been obtained earlier from rank-level duality
\FSch, but now for the first time
    they can be treated on an equal footing for all WZW-models: they
can be identified with the modular matrices $S$ of the `orbit Lie algebras'
      that are associated to
the Dynkin diagram automorphisms induced by the currents $J$.
Although the fixed point resolution matrices $S^J$
can in principle be extracted from \FSSc\ or \FSch, we believe it is
worthwhile to present the result in a more accessible  way.

\noindent
The term `conjecture' is used in the first item because
the conditions we solve are necessary, but not sufficient.
An important condition that in the general case is not easy to impose
is that the new matrix $\tilde S$
must yield sensible fusion rule coefficients when substituted in
Verlinde's formula.
(Note that a rigorous proof of the conjecture would require in
particular an explicit construction of the extended chiral algebra,
as well as the proof that
it gives rise to a reasonable conformal field theory.)

However, there are several reasons
why we believe our solution is the correct one, namely:
\item\dash The solution is {\it mathematically natural} in the sense
that a very simple closed formula can be given that applies to all cases.
\item\dash It has been checked by explicit computation to give
non-negative integer fusion coefficients for all types of simple
and many semi-simple algebras (of course, such checks have been
done only for a limited range of ranks and levels).
\item\dash It can be {\it derived} rigorously as the matrix $\tilde
S$ that describes the transformation of the characters of
diagonal coset models.

Our results also allow the computation of the matrix $\tilde S$ for
most coset models. The modular properties
of coset models $G/H$ can be described in terms  of a formal
tensor product  of the $G$-theory with the complement of the
$H$-theory (the complement of a conformal field theory has by
definition a complex conjugate representation of the modular group).
One gets a matrix $S_G \otimes S_H^*$ that acts on  the
branching functions. In many cases some of the branching functions
vanish, while others have to be identified with each other,
and correspond to  a single primary field
in the theory. This is known as {\it field identification}.
Field identification can be formulated
-- as far as modular transformation properties are concerned -- in
terms of a simple current extension of this tensor product,
except in a few rare cases (the so-called `maverick' cosets \DuJo).
Hence the computation of the matrix $\tilde S$ of coset
models is technically identical to the computation for a suitably
chosen integer spin simple current invariant
so that our conjecture regarding fixed point
resolution for $\tilde S$ covers this case as well.

However, there is an essential difference in the interpretation  and
computation of the fixed point characters.
In an integer spin modular invariant each of the representations
originating from a fixed point has the same character
with respect to the chiral algebra of the unextended theory,
namely the one appearing within the absolute value symbol in \Pfunct.
On the other hand in coset models the latter character is to
be interpreted as the {\it sum} of  $N$ characters that may be (and
in general are) distinct as functions of $\tau$. Hence the degeneracy
is lifted, and we can determine $\tilde S$ directly from the
transformation of the characters. All of this is useful only if one
is able to compute the coset characters, which
     for $N>1$
are not equal to the branching functions.
The differences between the branching functions and the
coset characters are called {\it character modifications}.

We have accomplished this for the diagonal coset models $G\times G/
G$, by realizing field identification on the entire Hilbert space, and
identifying the various eigenspaces of field identification on the
fixed points \FSSb. Having done this, we can prove that for diagonal
coset models the character modifications are equal to branching
functions of twining characters. Twining characters have been defined
in \FSSc\ and will be briefly described in section \TwineChap.
For our present purpose
all we need is the fact that in \FSSc\ the modular transformations
of these characters were obtained. This allowed us to derive
the modular transformations of the characters of diagonal coset
models. The
formula for $\tilde S$ we conjecture here is a generalization of the
one in \FSSb. The formula is not identical, since in the
general case a complication
arises that does not occur for diagonal coset models.

While in the case of coset conformal field theories and for
integer spin simple current invariants of WZW-theories the associated
orbit Lie algebras provide natural candidates for the matrices
$S^J$ that implement fixed point resolution, it is not clear whether
analogous
data are available for arbitrary rational conformal field theories.
However, we
expect that the matrices that describe the transformation of the
one-point functions of the simple currents on the torus will do the
job. Note that it follows on quite general reasons that these one-point
functions have good modular transformation properties \Mooresei\ and
are non-zero only for fixed points. The identification of the
matrices $S^J$ with the $S$-matrices
for torus one-point functions implies in particular
      the conjecture
that in the case of WZW-models the modular transformation properties
of these one-point functions are described by the orbit Lie algebras.
Apart from being conceptually elegant, this has the practical
advantage that an explicit
       closed
formula for the matrices $S^J$ can be given,
namely the Kac-Peterson \KaPe\ formula of the orbit Lie algebra.

The organization of this paper is as follows. In the next
       section
we formulate the conditions we impose on the solution. They consist
of six conditions that are beyond question, plus two additional
ones that should be considered
as working hypotheses.
In section 3 we discuss what can be deduced about $\tilde S$ using
only the six unquestionable conditions.

In
        section
4 we perform a Fourier transformation on the labels
of the resolved fixed points. If one imposes the two additional
conditions, this suggests an
{\it ansatz} for the matrix $\tilde S$ in the general case, and leads
naturally to a definition of the quantities $S^J$. This {\it ansatz} is
an additional assumption, and for this reason we do not claim to have
found the most general solution satisfying all conditions.
The characterization of the primary fields of the extended theory and the
 formula for $S$, given by equation \XX, are the main results of this paper.
It can be shown to satisfy all conditions given certain properties
of the matrices $S^J$. This proof is independent of the heuristic
arguments leading us to \XX, and is briefly summarized
in section 5. Readers who are only interested in the result may
therefore in fact skip sections 3 and 4.

In section 6 we briefly review the concepts of twining characters and
orbit Lie algebras and apply our formalism to WZW-models. Realizing that
the WZW-model based on $G=\tilde G/{\cal Z}$, where $\tilde G$
is the universal covering Lie group of $G$ and ${\cal Z}$ a subgroup
of the center of $\tilde G$, is described by the corresponding
simple current invariant, this leads in particular to a conjecture
for the $S$-matrix of WZW-models based
on  non-simply connected compact Lie groups
 (for the precise definition of these models see \fegk).

\chapter{Conditions}

With respect to the fusion product, the set of simple currents of
a conformal field theory forms a finite abelian group, known as the
center ${\cal C}$ of the theory. To any subgroup $\G \subseteq {\cal
C}$ of mutually local integral spin simple currents one can associate
a modular invariant partition function in which the
chiral algebra is extended by this set of currents.
  (An explicit expression for the partition function will be given in
\FormOne.)
Our goal is to write down for any such extended theory a pair of
matrices $\tilde S$ and $\tilde T$, which must satisfy the following
requirements:
$$ \eqalign{
{\rm [I]}  & \qquad \hbox{$\tilde S$ and $\tilde T$ act `correctly'
on the characters}.\cr
{\rm [II]} & \qquad \hbox{$\tilde S$ is symmetric}.\cr
{\rm [III]}& \qquad \hbox{$\tilde S$ is unitary}.\cr
{\rm [IV]} & \qquad \hbox{$\tilde S^2=\tilde C$}.\cr
{\rm [V]}  & \qquad \hbox{$\tilde S$ and $\tilde T$ satisfy
$(\tilde S\tilde T)^3=\tilde S^2$}.\cr } \eqn\Condition $$
Here $\tilde C$ is a matrix with entries $0$ or $1$ satisfying
$\tilde C^2=1$,
\ie\ a permutation of order 2, which furthermore acts trivially on
the identity. The
characters of the theory are linear combinations of characters of
the unextended theory. This gives us some information about their
modular transformations in terms of the matrices
$S$ and $T$ of the unextended theory. The meaning of the first
condition is that the matrices  $\tilde S$ and $\tilde T$
must reproduce this knowledge.
This is the only condition that relates $\tilde S$ to the matrix $S$
of the unextended theory. The matrix $\tilde T$ follows in a
straightforward way
from $T$ using condition [I], and therefore we do not specify any
explicit conditions for it. As usual, it is a unitary diagonal matrix.

Although we do not impose a general integrality condition
on the fusion rules derived from $\tilde S$, we make an exception for
certain simple current fusion rules, because they
are of special importance to us, and {\it can} be analyzed.
Suppose we are given a simple current $J$ in the unextended
theory that is `local' (i.e., has zero monodromy charge) with respect
to all currents in \G, so
that its orbit is an allowed field in the extended theory, i.e.\ $J$
is not `projected out' by the extension. We claim that this orbit
gives rise to a simple current in the extended
theory\rlap.\foot{Note, however, that fixed point
resolution might introduce additional simple currents that are not
related to simple currents of the unextended theory.}
Note that neither the identity primary field nor the simple current $J$
are fixed by the currents in $\G$. It is then easy to see that
the $S$-matrix of the extended theory satisfies $\tilde S_{0,J}=
\tilde S_{0,0}$. It follows that -- if indeed $\tilde S$ leads to
correct fusion rules -- the primary field $J$ in the extended theory
has quantum dimension 1 and hence again is a simple current. Therefore we
require
  $$ {\rm [VI]} \qquad \hbox{$ \tilde {\cal N}_{J,b}^{~~~c}
  = \delta_{Jb,c}.$~~~~~~~~~~~~~~~~~~~~~~~ } \eqn\ConditionOneb $$
Here $\tilde {\cal N}$ are the fusion coefficients obtained from
$\tilde S$ via the Verlinde \EVeA\ formula, and $Jb\equiv J\tims b$
is another primary field in the extended theory, obtained as the
fusion product of $J$ and $b$.
The fusion coefficients are finite since $\tilde S_{0,n}\not=0$
after fixed point resolution.

Conditions [I] -- [VI] are clearly necessary. It will be helpful to
impose two additional working hypotheses, namely
  $$ \eqalign{
  {\rm [VII]}& \qquad \hbox{Consistency of successive extensions}.\cr
  {\rm [VIII]}& \qquad \hbox{Fixed Point Homogeneity}.\cr
  } \eqn\ConditionTwo $$
Condition [VII] applies when there are several distinct paths to the
final result. This is the case if
there exist several distinct chains of subgroups of the form
  $$ \G\equiv\H_0 \supset \H_1\supset \ldots\supset
  \H_n \equiv \{\one\} \,, \eqn\Chain$$
which is possible if the order of $\G$ is not prime. If we have a
general formula that can deal with any extension, in particular it will
give a result for each such chain, when the extensions are performed
successively. Condition [VII] states that the answer should not
depend on the specific chain chosen. Condition
[VIII] means that in the final result the resolved fixed points
coming from the same primary field $a$ are
indistinguishable in as far as their modular transformations and
fusion rules are concerned. This condition has to be handled with
some care; while for coset theories it does hold for $\tilde S$ and the
fusion rules (in all known cases),
it does {\it not} apply to the characters, for which one has to
include so-called character modifications.

An integer spin simple current
modular invariant has the general form
  $$ Z=\sum_{\scriptstyle{\rm orbits}\,a \atop \scriptstyle Q=0}  |
  \S_a  | \cdot  |\!\!\!\sum_{J \in \G/\S_a}\!\!\! \X_{J a} |^2 .
  \eqn\FormOne $$
Here $\G$ is a subgroup of the center whose elements have integer
spin; the first sum is over all $\G$-orbits of primary fields in the
theory with zero monodromy charge $Q$. The monodromy charges of primary
fields $a$ with respect to the simple current $J$ are defined as
the fractional parts
  $$ Q_J(a) = h(a) + h(J) - h(Ja) \bmod\,\Zbf $$
of combinations of conformal weights.
The $S$-matrix elements of fields on the same simple current orbit are
related by
  $$ S_{La,b} = \ee^{2\pi\ii Q_L(b)} S_{a,b}  \,. \eqn\CSR $$
The group $\S_a$ appearing in \FormOne, the {\it stabilizer} of $a$,
is the subgroup of $\G$ that acts trivially on the orbit $a$;
$\X_{a}$ is the character of the field $a$, and
$\X_{ Ja}$ is the character of the representation obtained when the
simple current $J$ acts on the orbit $a$. Since the center $\cal C$
is abelian, all elements of an orbit have the same stabilizer $\S_a$.
The representations in the orbit $a$ are called {\it fixed points} with
respect to the currents in the stabilizer. Below we will also use the notation
  $$\G_a:= \G / \S_a   \eqn\GSa $$
for the factor group of currents that acts non-trivially on $a$.

Implicit in the foregoing discussion is a choice of a representative
within each orbit. In the following $a,b,c,\ldots$ always refer to a
definite choice of orbit representatives. The primary fields in the
unextended  theory are
then obtained as $Ja$ with $J \in \G_a$. Quantities like $S$, $T$ and
the fusion coefficients
 ${\cal N}$
in the extended theory will be distinguished by a tilde.

On general grounds
one expects \MoSc\ that it should be possible to rewrite
  the invariant \FormOne\
as a standard diagonal one, \ie\ as
  $$ Z=\sum_{\alpha}    |\Xt_{\alpha} |^2 \ .  \eqn\FormTwo $$
The matrix $\tilde S$ acts on the new characters $\Xt_{\alpha}$.
The relation between \FormTwo\ and \FormOne\ is straightforward
if there are no fixed points, \ie\ if  $|\S_a|=1$ for all $a$. If
$|\S_a|=2$ or 3 there is only one possible interpretation, namely
that the orbit $a$ corresponds to precisely $|\S_a|$ representations
in the extended theory. The characters of those representations are
identical with respect to the unextended algebra. If $|\S_a|\geq 4$
there are as many interpretations as there are ways of writing
$|\S_a|$ as a sum of squares. Each such square can be
absorbed in the definition of an extended character
 $\tilde\X_{\alpha}$
rather than being interpreted as a multiplicity.

In general there is thus ample room for ambiguities. First of all,
even the number of primary fields is not evident. Furthermore, for each
possible choice of the spectrum (which fixes the matrix $\tilde T$)
there may exist more than one matrix $\tilde S$ that satisfies
\Condition\ and \ConditionOneb.

\chapter{Fixed point resolution: Generalities}

We will now examine the consequences of the first six
conditions. The other two will be discussed later.

\section{Condition [I]}

\noindent
Each character $\Xt_{\alpha}$ of the extended theory is in any case
a sum of characters of the original theory, which
 belong
to a definite orbit $a$. There may be more than one character of
the extended theory that belongs
to the same $\G$-orbit, so we need an extra label.
Let us write $|\S_{a}|$ as a sum of squares,
  $$ |\S_{a}| = \sum_{i} (m_{a,i})^2 \ , \eqn\SqDecomp$$
where $i$ labels the different primaries into which $a$ gets resolved
(if we
also impose condition  [VIII], then $m_{a,i}$
         has
to be independent of $i$).
Corresponding to this definition we have
  $$ \Xt_{a,i}= m_{a,i} \sum_{J\in\G_a}\X^{}_{Ja} \,\eqn\ChiSplit $$
so that $\sum_i | \Xt_{a,i} |^2 = |\S_{a}|\cdot|\sum_{J\in\G_a}\,
\X_{Ja}|^2 $. Clearly $\tilde T_{(a,i),(a,i)}=T_{a,a}$
independent of $i$. For $\tilde S$ we find
  $$  \Xt_{a,i}(-{1\over\tau})= m_{a,i} \sum_{J\in\G_a}
  \sum_b \sum_{K\in\G_b} S_{Ja,Kb} \,\X_{Kb}(\tau)\,. $$
Here
    and in the rest of this paper
we write only the dependence on $\tau$, but there might be
additional variables (for example Cartan angles in affine Lie algebra
characters). This may in fact be necessary to resolve ambiguities
    in the unextended theory.
The last two sums form together a sum over all fields in the theory,
but because of the sum on $J$ only
those fields $Kb$ contribute that have zero monodromy charge with
respect to all currents in $\G$. We denote this as $Q_{\G}=0$. For
fields $Kb$ with $Q_{\G}(Kb)=0$ the matrix element
$S_{Ja,Kb}$ is in fact independent of $J$, so we get
  $$ \Xt_{a,i}(-{1\over\tau})= m_{a,i}\, | \G_{a} |
  \sum_{b}\sum_{K \in \G_b} S_{a,b}\, \X_{Kb}(\tau) \ ,$$
where we have also used that as a consequence of
$Q_{\G}(a)=0$ we have $S_{a,Kb}=S_{a,b}$.
Now we are faced with the problem that  in general there is more
than one character associated with the orbit whose representative is
$b$. Hence we may write
  $$ \sum_{K \in \G_b}  \X_{Kb}(\tau) ={1\over N(\eta,b)} \sum_j
  \eta_{b,j} \tilde\X_{b,j}\,, $$
where, in order to satisfy \ChiSplit,
  $$ N(\eta,b) = \sum_j \eta_{b,j} m_{b,j} \ , $$
and $\eta_{b,j}$ is a set of coefficients that is present for any $b$
that splits into more than a single representation. We find thus the
following formula for $\tilde S$:
  $$ \tilde S_{(a,i),(b,j)}=m_{a,i}\,|\G_a|\,S_{a,b}\, {1\over
  N(\eta,b)}\, \eta_{b,j} + \Delta_{(a,i),(b,j)} \,. $$
Here $\Delta_{(a,i),(b,j)}$ is a possible extra term whose presence
cannot be inferred from the previous arguments, because of possible
degeneracies in the
set of characters. (We are assuming here that the set of
(generalized) characters of the unextended theory is linearly
independent, and we will in any case only  consider
degeneracies that were introduced by the fixed point
resolution.) These degeneracies allow for an additional term,
provided it satisfies
  $$\sum_j \Delta_{(a,i),(b,j)}\, m_{b,j} = 0 \ .\eqn\GammaCond$$

\section{Condition [II]}

\noindent
Now we impose the condition that $\tilde S$ must be symmetric.
Multiplying this condition with $m_{a,i}$ and summing over $i$ we get
  $$ |\G|\, {1\over N(\eta,b)}\,\eta_{b,j}\,S_{a,b}+\sum_i m_{a,i}\,
  \Delta_{(a,i),(b,j)}= m_{b,j}\,|\G_b|\, S_{a,b} \,. \eqn\mGS  $$
Now since $S$ is unitary,
for any $b$ there exists an $a$ such that $S_{a,b} \not = 0$. Let us
   fix $b$ and
pick one such $a$. Then \mGS\ can be solved for $\eta_{b,j}$:
  $${1\over N(\eta,b)}\,\eta_{b,j} = {|\G_{b}|\over |\G|}\,m_{b,j}-
  {\sum_i m_{a,i}\, \Delta_{(a,i),(b,j)} \over |\G_{b}|\, S_{a,b}}
  \,.  \eqn\EtaEqn$$
Note that the dependence of the last term on $a$ should cancel.
Substituting \EtaEqn\
into the formula for $\tilde S$ we get
  $$ \tilde S_{(a,i),(b,j)}=m_{a,i} m_{b,j}\, {|\G_a|\,
  |\G_b|\over|\G|} S_{a,b} + \Gamma_{(a,i),(b,j)} \,, \eqn\GenS $$
where the last term is equal to $\Delta$ plus the contribution from
the second term in \EtaEqn,
  $$  \Gamma_{(a,i),(b,j)} = \Delta_{(a,i),(b,j)} - m_{a,i}\,
  {|\G_a|\over|\G_b|}\,\sum_k m_{a,k}\Delta_{(a,k),(b,j)}  \,. $$
Note that $\Gamma$ satisfies a sum rule analogous to \GammaCond.
Furthermore symmetry of $\tilde S$ implies that $\Gamma$ must
be symmetric.

\section{Condition [III]}

\noindent
The remaining conditions involve a product $\P$ of two matrices,
either $\tilde S^2$, $\tilde S\tilde S^{\dagger}$
or $(\tilde S\tilde T)\tilde S$.
Note that $T$ is constant for fixed $a$ or $b$. As a consequence,
when we write such a product symbolically as $\P=\Pss+\Psg+\Pgs
+\Pgg$, then the cross-terms $\Psg$ and $\Pgs$ between the
two terms in \GenS\ always cancel due to condition \GammaCond.
For $\P=\tilde S \tilde S^{\dagger}$, the term $\Pss$ reads
$\Pss=\sum_{b,j}m_{a,i}m_{b,j}^2 m_{c,k}\,S_{a,b}\,S_{b,c}^{\dagger}
\, |\G_a| |\G_b|^2 |\G_c|\,/\,|\G|^2$.
The sum over $j$ can be done using \SqDecomp:
  $$ \Pss= \sum_b m_{a,i}m_{c,k}\, {|\G_a|\, |\G_b|\, |\G_c|\over
  |\G|}\, S_{a,b}^{} S_{b,c}^{\dagger}     
  = \sum_b \sum_{J,K} m_{a,i} m_{c,k}\, { |\G_c|\over |\G|}\,
  S_{Ja,Kb}^{}\, S_{Kb,c}^{\dagger} \,. $$
Here we have traded the factor $| \G_{a} ||\G_b| $
for  a sum over the orbits of $a$ and $b$. Each term in these orbits
gives the same contribution. The
sum on $b$ is over all orbit representatives of $Q_{\G}=0$ orbits.
It can be extended to a sum over {\it all} orbits because the
contributions of the  $Q_{\G}\not =0$ orbits cancel among each other
owing to the sum on $J$. Together with the sum on $K$ we now have a sum
over all primary fields in the unextended theory, and we can use
unitarity (respectively $S^2=C$, or $STS=T^{-1} S  T^{-1}$) in the
unextended theory.
{\small{Now we get
  $$ \Pss=  \sum_J m_{a,i} m_{c,k}\, { |\G_c|\over |\G|}\,
  \delta_{Ja,c} = m_{a,i} m_{a,k}\, { |\G_a|\over |\G|}\, \delta_{a,c}
  \,. $$
Note that in the first term $\delta_{Ja,c}$ is a Kronecker
symbol in the unextended theory. It implies that
$a$ and $c$ must be representatives from the
same orbit, and hence must be the same. Together with the
contribution $\Pgg=\sum_{b,j} \Gamma_{(a,i),(b,j)}^{}
\Gamma_{(b,j),(c,k)}^{\dagger}$ we should get
$\delta_{ac}\delta_{ik}$.} }
Requiring unitarity of $\tilde S$ leads then to the condition
  $$\sum_{b,j} \Gamma_{(a,i),(b,j)}^{}\Gamma_{(b,j),(c,k)}^{\dagger}=
  \delta_{ac} \left(\delta_{ik} - {m_{a,i}\,m_{a,k} \over |\S_a|}
  \right) \,. $$
Note that the right hand side is a projection operator,
  $$ P^a_{ik} \equiv \delta_{ik} - { m_{a,i}\, m_{a,k}  \over |\S_a|}
  \,. $$
A special case of this result was already obtained in \ScYe, but
there all multiplicities $m_{a,i}$ were assumed to be equal to 1.

\section{Condition [IV]}

\noindent
The computation for $\tilde S^2$ yields in a similar way the relation
  $$\sum_{b,j} \Gamma_{(a,i),(b,j)}\Gamma_{(b,j),(c,k)}=
  \tilde C_{(a,i),(c,k)} - {m_{a,i}\,m_{c,k} \over |\S_c|} \sum _J
  C_{Ja,c}\,, $$
where $\tilde C_{(a,i),(c,k)}$ is the charge conjugation matrix
of the extended theory and $C_{a,c}$ that of the unextended one.
The sum in the second term can only contribute if $a$ and $c$
are representatives of conjugate orbits, and in that case it
contributes 1, and otherwise 0. We may thus introduce a
matrix $\hat C_{a,c}$ on orbit representatives which is 1 if $a$
is conjugate to some field on the ${\G}$-orbit of $c$, and 0
 otherwise. Then we get
  $$\sum_{b,j} \Gamma_{(a,i),(b,j)}\Gamma_{(b,j),(c,k)}= \tilde
  C_{(a,i),(c,k)}
  - { m_{a,i}\, m_{c,k}  \over |\S_c|}\, \hat C_{ac} \,. $$
Using the sum rule analogous to \GammaCond\ that is valid for the
matrix $\Gamma$, we conclude that
$ \sum_k \tilde C_{(a,i),(c,k)}\,m_{c,k} = m_{a,i}\, \hat C_{a,c}.$
This implies that $\tilde C_{(a,i),(c,k)}$ can only be non-zero
between orbits $a$ and $c$ with $\hat C_{a,c}=1$. Furthermore,
conjugate fields must have the same value of $m$. It follows
that the set of numbers $m_i$ must be identical on conjugate
 orbits. Hence we may write
  $$ \tilde C_{(a,i),(c,k)} = \hat C_{a,c}\, C^c_{i,k} \,,\eqn\GenSQ
  $$
where $C^c_{i,k}$ is a conjugation matrix that is
introduced by the fixed point resolution.
Because $\tilde C$ and $\hat C$ are symmetric, the
  matrices
$C^c$ must satisfy
  $$ C^c= (C^{c^*})^{\rm T}\  $$
if $c^*$ is the conjugate of $c$. The final result is therefore
  $$\sum_{b,j} \Gamma_{(a,i),(b,j)}\Gamma_{(b,j),(c,k)}=\hat C_{a,c}
  \sum_l C^c_{i,l}\,P^c_{l,k} \,. $$

\section{Condition [V]}

\noindent
   Condition [V] is most conveniently dealt with in the equivalent form
   $\tilde S \tilde T \tilde S = \tilde T^{-1} \tilde S \tilde T^{-1}$.
For $\tilde S\tilde T\tilde S$ we find
  $$\sum_{b,j} \Gamma_{(a,i),(b,j)}\,\tilde
T_{b,b}\,\Gamma_{(b,j),(c,k)}=
  (\tilde T^{-1}\tilde S\tilde T^{-1})_{(a,i),(c,k)}
  - \sum _J m_{a,i} m_{c,k}\, {|\G_c| \over |\G|}\,
  T^{-1}_{Ja,Ja}\, S_{Ja,c}\, T^{-1}_{c,c} \,. $$
The sum on $J$ just yields a factor $|\G_a|$, and then the
last term cancels the first contribution from $\tilde S$. The result
is  $$\sum_{b,j} \Gamma_{(a,i),(b,j)}\,\tilde
T_{b,b}\,\Gamma_{(b,j),(c,k)}=
  (\tilde T^{-1}\Gamma\tilde T^{-1})_{(a,i),(c,k)} \,. $$

\section{Some remarks on fusion rules}

\noindent
Although it seems to be quite difficult to examine the
fusion rules in general, we can discuss the case that one of the
three fields
is {\it not} a fixed point (if even fewer fields are fixed points,
the discussion is completely straightforward).
Note that the formula $\sum_i m_{a,i}
\Gamma_{(a,i),(b,j)} = 0 $ implies that $\Gamma=0$ if a field
is `resolved' into only one primary field of the extended theory. In
particular, there are no correction matrices $\Gamma$ for fields
that are not fixed points.

{\small{
The calculation is done in the following way. From the Verlinde
formula we get
  $$\eqalign{
  \tilde {\cal N}_{a,(b,j)}^{\phantom{a,(b,j)}\!\!(c,k)} &=
  \sum_{d,l}{ m_a m_{d,l} {| \G_a | |\G_d| \over |\G| } S_{a,d}
\over
  m_0 m_{d,l} {|\G_0| |\G_d|\over |\G|} S_{0,d} }\cr &\times
  \left[ m_{b,j} m_{d,l} {| \G_b | |\G_d| \over |\G| } S_{b,d}
  + \Gamma_{(b,j),(d,l)} \right]
  \left[ m_{c,k} m_{d,l} {| \G_c | |\G_d| \over |\G| } S_{c,d}
  + \Gamma_{(c,k),(d,l)} \right]^* \,. \cr }$$
In the first factor $m_a=m_0=1$, $\G_a=\G_0=\G$, so that we get
  $$ \tilde {\cal N}_{a,(b,j)}^{\phantom{a,(b,j)}\!\!(c,k)} =
  \sum_{d,l}{S_{a,d}
  \over S_{0,d} } \left[ m_{b,j} m_{d,l} {|\G_b| |\G_d| \over |\G| }
  S_{b,d}
  + \Gamma_{(b,j),(d,l)} \right] \times \left[ m_{c,k} m_{d,l}
  {|\G_c| |\G_d| \over |\G| } S_{c,d} + \Gamma_{(c,k),(d,l)}
  \right]^* \,. $$
Because of the sum rule for $\Gamma$ the cross-terms vanish. In the
$SS^*$-term
we use $\sum_l m_{d,l}^2 = |\G|/|\G_d|$, and we convert the remaining
factor $\G_d$ to a sum over the $d$ orbit. The factor $\G_b$ can
likewise
be converted to a sum over the $b$ orbit. If we sum over the full
$b$ orbit, we may extend the sum over $d$ to include also fields that
have non-zero charge with respect to the extended algebra, since for
such fields
  $$ \sum_{J \in \G_b} S_{Jb,d} = {1\over |\S_b| } \sum_{J \in \G}
  S_{Jb, d}
  = {1\over |\S_b| } \sum_{J \in \G} \ee^{2\pi \ii Q_J(d) }S_{b,d} =
  0 \,.  \eqn\Cancel $$
(This argument was already used before.)
The fusion coefficients are thus
  $$ \tilde {\cal N}_{a,(b,j)}^{\phantom{a,(b,j)}\!\!\!(c,k)}
  = {|\G_c|\over|\G|}\, m_{b,j}m_{c,k} \sum_{J\in \G_b}
  {\cal N}_{a,Jb}^
    {\phantom{a,Jb}c} + \sum_{d,l} {S_{a,d} \over S_{0,d} }\,
  \Gamma_{(b,j),(d,l)}\Gamma_{(c,k),(d,l)}^*\,, $$
which is equivalent to the more symmetric result given below. }}

We obtain the following formula for the fusion coefficients:
  $$  \tilde {\cal N}_{a,(b,j)}^{\phantom{a,(b,j)}\!\!\!(c,k)} =
  {|\G_b|\, |\G_c|\over |\G|^2}\, m_{b,j} m_{c,k}\,
  \sum_{J\in \G} {\cal N}_{a,Jb}^{\phantom{a,Jb}\!c} + \sum_{d,l}
  {S_{a,d} \over S_{0,d} }\,
\Gamma_{(b,j),(d,l)}\Gamma_{(c,k),(d,l)}^* \,. $$
In principle the requirement that the coefficient should be a positive
integer imposes restrictions on $\Gamma$, but these conditions do not
look particularly useful. This is even more true for the fusion
of three resolved fixed points. A few things can be learned, though.
Multiplying with $m_{c,k}$ and summing over $k$ we get
  $$ \sum_k \tilde {\cal N}_{a,(b,j)}^{\phantom{a,(b,j)}\!\!\!(c,k)}
  m_{c,k} = m_{b,j} \sum_{J \in \G_b} {\cal N}_{a,Jb}^{~~~~c} \,. $$
This has a few implications. If $a \tims b$ contains terms in the
orbit of $c$, then $\tilde {\cal
N}_{a,(b,j)}^{\phantom{a,(b,j)}\!\!\!(c,k)}$
cannot be zero for all $k$. Furthermore, if $a\tims b$ does not
yield any contribution in the $c$ orbit, then
$\tilde {\cal N}_{a,(b,j)}^{\phantom{a,(b,j)}\!\!\!(c,k)}=0$. Thus
the new fusion rules must respect the orbit-orbit maps of the original
fusion rules, although the distribution of fields may be non-trivial.

We can in fact say more. If $n= \sum_{J\in\G_b} {\cal N}_
{a,Jb}^{~~~~c}=1$ and
 $|\G_b| = |\G_c|$,
then it can be shown that the vector $\vec m_c$ is a permutation of
$\vec m_b$. Since this is not a surprising result, we omit the
details of the proof.
{\small{The proof goes as follows.
We have seen that conjugate fields have the same value of $m_i$,
\ie\ $m_{a,i}=m_{a^*,i}$, where we use the same labelling for fields
and their conjugates. Now define
  $$ n= \sum_{J \in \G_b} N_{a,Jb}^{~~~~c} \,. $$
Then
  $$ \sum_k  \tilde N_{a,(b,j)}^{\phantom{a,(b,j)}\!\!\!(c,k)}
  m_{c,k} = n\, m_{b,j} \,, \eqn\mbj$$
and also
  $$ \sum_i  \tilde N_{a,(c^*,k)}^{\phantom{a,(b,j)}\!(b^*,i)}
  m_{b^*,i} = n^*\, m_{c^*,k} \,,\eqn\mck $$
with
  $$ n^* =  \sum_{J \in \G_{c^*}}
  N_{a,Jc^*}^{~~~~~b^*}\,.\eqn\nstarOne $$
Because of the sum over $J$ only fields with non-zero
$J$-charge contribute to the sum in Verlinde's formula.
This allows us to average over the orbit of $a$:
  $$ n^* = {1\over |\G_a|} \sum_{K \in \G_a}\sum_{J \in \G_{c^*}}
  N_{Ka,Jc^*}^{~~~~~~~b^*}\ , \eqn\nstar $$
since all terms in the sum over $K$ give the same contribution. For
the same reason, in the presence of the sum over $K$ all terms
in the sum over $J$ give the same contribution.
We may thus replace the sum over $J$ by a factor
$|\G_c|$. Now we may include a summation over the $b$ orbit,
compensated by a factor ${1/ | \G_b| }$, and finally remove the
sum over the $a$ orbit again. Also noting that
  $$ N_{ab}^{~~~c} = N_{ac^*}^{~~~b^*} \,, $$
the result becomes
  $$ n^* = {|\G_c| \over |\G_b|} \sum_{J \in \G_{b^*}}
  N_{a,c^*}^{~~~Jb^*}=
  {|\G_c| \over |\G_b|} \sum_{J^* \in \G_{b^*}} N_{a,c^*}^{~~~(Jb)^*}
  = {|\G_c| \over |\G_b|}\, n\,.$$
Hence we have
  $$ m_{c,k}= {|\G_b|\over n |\G_c|}
  \sum_i \tilde  N_{a,(c^*,k)}^{\phantom{a,(b,j)}\!\!(b^*,i)}\,
  m_{b,i}= {|\G_b|\over n |\G_c|}
  \sum_i \tilde  N_{a,(b,i)}^{\phantom{a,(b,j)}\!\!\!(c,k)}\, m_{b,i}
  \,. $$
Next we substitute the relation \mbj:
  $$ m_{c,k}=m_{c^*,k}={|\G_b|\over n^2|\G_c|}
  \sum_i  N_{a,(b,i)}^{\phantom{a,(b,j)}\!\!\!(c,k)}
  \sum_l  N_{a,(b,i)}^{\phantom{a,(b,i)}\!\!\!(c,l)}\, m_{c,l} \,. $$
Consider now the special that $n=1$ and
$|\G_b|=|\G_c|$. Then the equation is of the form
  $$ \vec m_c = M M^T \vec m_c \ ,$$
where $M_{ik}=N_{a,(b,i)}^{~~~~(c,k)}$. $M$ is non-negative definite,
and furthermore
  $$ (MM^T)_{kk}=\sum_i
  N_{a,(b,i)}^{~~~~(c,k)}N_{a,(b,i)}^{~~~~(c,k)}  \geq 1 \,, $$
because for fixed $k$ the fusion coefficient does not vanish for all
$i$, since $n\not=0$. Hence we may write $ MM^T = 1 + X $,
where $X$ is a non-negative matrix that must satisfy
$X\vec m_c=0$. But this implies that $X=0$, since
all components of the vector $\vec m_c$ are positive.
Hence $MM^T=1$. This implies that $M$ is a permutation, and therefore
the vector $m_{b,j}$ is a permutation of $m_{c,k}$.}}%
The condition that $n=1$ and $|\G_b|=|\G_c|$ is in particular
satisfied if $a$ is a simple current, but in general this is
by no means the only possibility.
If there is any non-fixed point field $a$ that maps  orbit $b$ to
$c$ with multiplicity 1, then orbit $b$ and $c$
must have the same decomposition vector $\vec m$ (if
 $|\G_b|=|\G_c|$).
The result indicates that fixed points which have the same stabilizer should
also possess the same decomposition $\vec m$; it is difficult to imagine
how a different decomposition could still provide a solution to all
constraints.

\section{Condition [VI]}

\noindent
Suppose there is a simple current $L$ in the theory which is local
with respect to all currents by which we extend the algebra. Then
according to the remarks before equation \ConditionOneb\
$L$ will again be a simple current in the extended theory.
(Note that, as before, $L$ stands for a definite
representative of the $\G$-orbit of the additional currents.)
Hence $L$ must act as a simple current on the resolved fixed points.
Thus if in the unextended theory
  $$ L \tims a = b \,, $$
then as seen above we have in the extended theory
  $$ L \tims \ai = \bj $$
for some $j$.  Hence we have both
  $$ Q_L(a)=h(a)+h(L)-h(b) \mod 1\hbox{~~~~~~~~~}
  \eqn\ChargeDefinition$$
and
  $$ \hbox{~~~~~~} \tilde Q_{L}(\ai)=h(\ai)+h(L)-h(\bj) \mod 1 \,. $$
Since the respective conformal weights are the same up to integers,
we see that on all fields $Q_L=\tilde Q_{L}$. Note that $L$ was
assumed to be local with respect to $\G$, so that $h(L)$ is a
constant (modulo integers) on the $\G$-orbit of $L$, and hence
the notation makes sense.

 To relate the matrix element
$\Gamma_{(a,i),(c,k)}$ to
$\Gamma_{(b,j),(c,k)}$, we recall the relation \CSR\
that a simple current $L$ imposes on the $S$-matrix elements.
Combining this formula with the analogous relation
  $$ \tilde S_{L(a,i),(b,j)}=\ee^{2\pi\ii Q_L(b)} \tilde
  S_{(a,i),(b,j)}  \eqn\CSREF $$
for $\tilde S$, we obtain an analogous relation for $\Gamma$:
{\small{The simple current structure of $\tilde S$ implies in
particular that
  $$\eqalign{ \tilde S_{L (a,i),(c,k)} &= \ee^{2\pi\ii\tilde
  Q_{L}(\ck)}
  \tilde S_{(a,i),(c,k)} = \ee^{2\pi\ii Q_L(c)}\tilde S_{(a,i),(c,k)}
  \cr&
  = \ee^{2\pi \ii Q_{L}(c)} \left[m_{a,i} m_{c,k}\, {|\G_a| |\G_c
  |\over |\G|}\,
  S_{a,c} + \Gamma_{(a,i),(c,k)}\right]\,. }\eqn\JOne $$
On the other hand,
  $$\eqalign{ \tilde S_{L(a,i),(c,k)}=S_{(b,j),(c,k)}&=
  m_{b,j}m_{c,k}\,{|\G_b| |\G_c|\over |\G|}\, S_{b,c} +
  \Gamma_{(b,j),(c,k)}\cr
  &= m_{b,j} m_{c,k}{|\G_b||\G_c|\over |\G|}\,
  S_{La,c}+\Gamma_{L(a,i),(c,k)}\cr
  &= \ee^{2\pi \ii Q_{L}(c)}m_{b,j} m_{c,k}\, {|\G_b| |\G_c|\over
  |\G|}\, S_{a,c} + \Gamma_{L(a,i),(c,k)}\cr
  &= \ee^{2\pi \ii Q_L(c)}m_{a,i} m_{c,k}\, {|\G_a| |\G_c|\over
  |\G|}\,  S_{a,c} + \Gamma_{L(a,i),(c,k)} \,.\cr} \eqn\JTwo $$
If $L \tims a = b $ it is easy to see that $\G_a \sim \G_b$, since
the simple current action is abelian. It follows from the arguments
given above that
$m_{b,j}=m_{a,i}$ if $\bi$ is the simple current image of $\ai$. This
explains the last step. Comparing \JOne\ and \JTwo\ we get}}%
  $$ \Gamma_{L(a,i),(c,k)}= \ee^{2\pi \ii Q_L(c)}\,
  \Gamma_{(a,i),(c,k)} \,. \eqn\SCGamma$$

\chapter{Fourier decomposition}

Suppose we consider an arbitrary fixed point resolution, where
a fixed point $a$ is split into $M_a$ primary fields. From now
on we will impose the homogeneity condition [VIII], and therefore in
particular we will only consider the case that $a$ is split into $M_a$
primary fields with identical multiplicity factors $m_{a,i}=m_a$. Then
  $$  |\S_a|=(m_a)^2 M_a \,. \eqn\smmm $$
Suppose by some as yet unspecified procedure we obtain a matrix
$\tilde S_{(a,i),(b,j)}$ satisfying all the requirements listed in
section 2.

\section{Group characters}

\noindent
For each fixed point choose an abelian discrete group $\M_a$
with as many characters as there are resolved fields, \ie\
 $|\M_a|=M_a$. Later we will identify this group as a subgroup of the
stabilizer, but for the moment there is no need to be specific.
An important role will be played by the group characters
$\Psi^a_i$, $i=1,2,...\,,M_a$, of $\M_a$. The characters are
a complete set of complex functions on the group satisfying
  $$ \eqalign{\Psi^a_i(g)\Psi^a_i(h)&=\Psi^a_i(gh) \,,\cr
  \Psi^a_i(g^{-1}) &= \Psi^a_i(g)^*  \,, \cr
  \Psi^a_i(\one)&=1  \,,\cr}\eqn\Inv $$
for all $g,h \in \M_a$ ($\one$ denotes the unit element of $\M_a$).
For these characters the orthogonality and completeness relations
  $$ \sum_i \Psi^a_i(g) \Psi^a_i(h)^*= M_a\, \delta_{gh} \,, \qquad
  \sum_g \Psi^a_i(g) \Psi^a_j(g)^*= M_a\, \delta_{ij} \eqn\Compl $$
hold. For cyclic groups $\Zbf_N$ we will label the elements by
integers $0\leq g<N$; the characters read
  $$  \Psi^a_{\ell}(g)= \ee^{2\pi \ii\,\ell g/N}\qquad{\rm for}\quad
  0 \leq \ell < N \,. $$
The groups $\M_a$ are chosen isomorphic on conjugate $\G$-orbits, as
well as on $\G$-orbits connected by any additional simple currents.
This is
possible since we have seen that the decomposition vector $\vec m$ is
preserved by charge conjugation and simple current maps.

We define the Fourier components of $\tilde S$ with respect to the
groups $\M_a$ as
  $$ S^{g,h}_{a,b} := {1\over \sqrt{M_a  M_b } }
  \sum_i \sum_j \Psi^a_i(g)^*\, \Psi^b_j(h)\, \tilde S_{(a,i),(b,j)}
  \,. $$
The inverse of this transformation is
  $$ \tilde S_{(a,i),(b,j)}={1\over \sqrt{M_a M_b} }\sum_g\sum_h
  \Psi^a_i(g)\, \Psi^b_j(h)^*\, S^{g,h}_{a,b} \,. \eqn\Fourier $$

Now we will examine the implication of conditions [I] -- [V] in terms
of the Fourier components.

\section{Condition [I]}

\noindent
Because of this condition some of the elements $S^{g,h}_{a,b}$ are
already known. According to the general expression \GenS\
for fixed point resolution, we have
  $$ \tilde S_{(a,i),(b,j)}=m_am_b\, {|\G|\over|\S_a|\, |\S_b|}\,
  S_{a,b} + \Gamma_{(a,i),(b,j)} \,. \eqn\GeNS $$
Using \GeNS, we can compute $ S^{g,h}$ for $g=\one$ (or $h=\one$).
The characters obey $\Psi^a_i(\one)=1$ for all $i$. Using also
  $$ \sum_i m_{a,i}\, \Gamma_{(a,i),(b,j)} = 0 $$
and the fact that the multiplicities are by assumption independent of
$i$, it follows that in this case
$\Gamma$ does not contribute. The only contribution is thus
  $$  S^{\one,h}_{a,b} = {1\over \sqrt{M_a   M_b  } }
  \sum_i\sum_j \Psi^b_j(h)\, {|\G|\,m_am_b\over |\S_a|\,|\S_b|}\,
  S_{a,b} \,.$$
Because of the orthogonality relation of the characters,
this vanishes unless $h=\one$.
For $h=\one$ the sums over $i$ and $j$ yield $ M_a M_b$, and the
result is
  $$ S^{\one,h}_{a,b}= S^{h,\one}_{a,b}
  ={|\G|\,m_a m_b\sqrt{M_a M_b}\over| \S_a|\, |\S_b| }\,
  \delta_{h,\one}\, S_{a,b}= {|\G|\over\sqrt{|\S_a|\, |\S_b|} }\,
  \delta_{h,\one}\, S_{a,b}\,. \eqn\Soneone$$

\section{Condition [II]}

\noindent
Symmetry of $\tilde S_{(a,i),(b,j)}$ implies
  $$ S^{h^{-1},g^{-1}}_{b,a}=S^{g,h}_{a,b} \,. $$

\section{Condition [III]}

\noindent
Unitarity of $\tilde S$ can be shown to be equivalent to
  $$ \sum_{h,b} S^{g,h}_{a,b}(S^{h^{-1},f}_{b,c})^*
 = \delta_{ac}\,\delta_{g,f^{-1}} \,. \eqn\SghUnit $$
{\small{The details are as follows:
  $$\eqalign{ \sum_{h,b} S^{g,h}_{a,b}(S^{h^{-1},f}_{b,c})^*
  &= \sum_b \sum_h{1\over \sqrt{M_a M_b} }
    \sum_{i,j,k,l} \Psi^a_i(g)^* \Psi^b_j(h) \tilde S_{(a,i),(b,j)}
    {1\over \sqrt{M_b M_c} } \Psi^b_k(h^{-1}) \Psi^c_l(f)^*
    \tilde S_{(b,k),(c,l)}^*\cr
  &=\sum_b{1\over \sqrt{M_a M_c} } \sum_{i,j,k,l}\Psi^a_i(g)^*
    \delta_{jk}
    \tilde S_{(a,i),(b,j)} \Psi^c_l(f)^* \tilde S_{(b,k),(c,l)}^*\cr
  &={1\over \sqrt{M_a M_c} }\sum_b \sum_{i,j,l}\Psi^a_i(g)^*
    \tilde S_{(a,i),(b,j)} \Psi^c_l(f)^* \tilde S_{(b,j),(c,l)}^*\cr
  &={1\over \sqrt{M_a M_c} } \sum_i \sum_l\Psi^a_i(g)^*
    \delta_{ac}\delta_{il} \Psi^c_l(f)^*
   ={1\over \sqrt{M_a M_c} } \sum_i   \Psi^a_i(g)^*
    \delta_{ac} \Psi^a_i(f)^*
   = \delta_{ac}\delta_{gf^{-1}}  \,.  \cr } $$
Proving the converse, namely that this relation implies unitarity
of $\tilde S$, is straightforward.}}

\vskip .2truecm

\section{Condition [IV]}

\noindent
Consider now the product $\tilde S^2$.  Using \GenSQ\ we find
  $$ \sum_{h,b} S^{g,h}_{a,b}S^{h,f}_{b,c} ={1\over \sqrt{M_a M_c} }
  \sum_i \sum_l\Psi^a_i(g)^* \Psi^c_l(f)\, \hat C_{a,c}\,C^c_{i,l}
  \,.   $$
{\small{The calculation is almost the same:
  $$\eqalign{ \sum_{h,b} S^{g,h}_{a,b}S^{h,f}_{b,c}
  &=\sum_b \sum_h{1\over \sqrt{M_a M_b} } \sum_{i,j} \Psi^a_i(g)^*
    \Psi^b_j(h)
    \tilde S_{(a,i),(b,j)} {1\over \sqrt{M_b M_c} }\sum_k \sum_l
    \Psi^b_k(h)^* \Psi^c_l(f)\tilde S_{(b,k),(c,l)}\cr
  &=\sum_b{1\over \sqrt{M_a M_c} } \sum_{i,j,k,l}\Psi^a_i(g)^*
    \delta_{jk}
    \tilde S_{(a,i),(b,j)} \Psi^c_l(f)\tilde S_{(b,k),(c,l)}\cr
  &={1\over \sqrt{M_a M_c} }\sum_b \sum_{i,j,l}\Psi^a_i(g)^*
    \tilde S_{(a,i),(b,j)} \Psi^c_l(f) \tilde S_{(b,j),(c,l)}\cr
  &={1\over \sqrt{M_a M_c} } \sum_i \sum_l\Psi^a_i(g)^* \Psi^c_l(f)
    \tilde C_{(a,i),(c,k)}
   ={1\over \sqrt{M_a M_c} } \sum_i \sum_l\Psi^a_i(g)^* \Psi^c_l(f)
    \hat C_{a,c}C^c_{i,l} \,.  \cr } $$}}%
This vanishes unless $c=a^*$. If $c=a^*$, then $M_a=M_c$, and we
may write the result as
  $$ \sum_{h,b} S^{g,h}_{a,b}S^{h,f}_{b,c}=\hat C_{a,c}\,{1\over M_c}
  \sum_{i,l} \Psi^c_i(g)^*\, \Psi^c_l(f)\, C^c_{i,l} \,. $$
If $C^c_{i,l}=\delta_{il}$ the result is $\hat C_{a,c}\delta_{gf}$.
Otherwise  $C^c_{i,l}$ defines a permutation of the labels $i$
of the characters. We may define
  $$ C^c_{g,f}:={1\over M_c} \sum_{i,l} \Psi^c_i(g)^*\,C^c_{i,l}\,
  \Psi^c_l(f) \,,$$
so that the result is
  $$ \sum_{h,b} S^{g,h}_{a,b}S^{h,f}_{b,c}=\hat C_{a,c}\, C^c_{g,f}
\,. $$
The matrix $C^c_{g,f}$ is unitary, but in general
it is not a permutation even though the Fourier transform $C^c_{i,l}$
is.

\section{Condition [V]}

\noindent
A completely analogous computation can be done for the
relation $(\tilde S\tilde T)^3=\tilde C$ of the modular group. The
result is
  $$ \sum_{h,b} S_{a,b}^{g,h}T_{b,b}^{}S^{h,f}_{b,c}= T^{-1}_{a,a}
  S^{g,f}_{a,c} T^{-1}_{c,c} \,. $$

\section{Condition [VI]}

\noindent
If the extended theory has a surviving simple current $L$ we have
  $$ \tilde S_{L(a,i),(b,j)}= \ee^{2\pi \ii Q_L(b)}
  \tilde S_{(a,i),(b,j)} \,. \eqn\SCR $$
The action of $L$ on the resolved fixed point moves the
orbit representative $a$ to $La$, and the label $i$ to $Li$. Here
$Li$ denotes some other label of the resolved fixed points of the
field $La$.
Expanding the left and right hand side of \SCR\ into Fourier modes,
we get $ \ee^{2\pi \ii Q_L(b)}S^{g,h}_{a,b} = {1\over M_a}
  \sum_{i,f} \Psi^a_i(g)^*\, \Psi^a_{\!Li}(f)\, S_{La,b}^{f,h}.$
Analogously as we did above for charge conjugation, we define a matrix
  $$ {\cal F}^a_{\!g,f}(L) :={1\over M_a} \sum_i
  \Psi^a_i(g)^*\, \Psi^a_{\!Li}(f) \,, $$
so that we can write the result as
  $$  \ee^{2\pi \ii Q_L(b)}S^{g,h}_{a,b} =
  \sum_{f} {\cal F}^a_{\!g,f}(L)\, S_{La,b}^{f,h} \,. \eqn\eSFS $$

\section{Condition [VII]: Successive extensions}

\noindent
Condition [VII] has several consequences. We will first compare
an extension in two steps with a complete extension, \ie\ in the
notation of \Chain\ we compare
  $$ \G \supset \H \supset \{\one\}\quad \hbox{with}\quad
  \G\supset\{\one\}\,.$$
For simplicity we consider only the case where  $p\equiv|\G|/|\H|$ is
prime, which by recursion includes all other cases anyway.

Performing the extension by $\H$, we obtain a modular matrix
$\tilde S_{(a,i),(b,j)}$, which can be described by Fourier components
$S^{g,h}_{a,b}$.
       By assumption
the extended theory has an integral spin simple current $L$ of prime
order $p$. When we further extend by this current $L$,
the stabilizer of any field $a$ remains either unchanged or is
enlarged from $\S_a^{\H}$ to $\S_a^{\G} \supset \S_a^{\H}$. If it
remains unchanged, then for $L\in \G \setminus \H$ we have
$L(a,i) = (b,j)$ with $a \not=b$, and hence $L$ has in any
case no fixed points; this situation requires no further discussion.
On the other hand,
if the stabilizer is enlarged, then $|\S^{\G}_a|=p\cdot|\S^{\H}_a|$.
Now two cases have to be distinguished:
\item\bullet Case A: \ $L (a,i) = (a,i)\,. $
\item\bullet Case B: \ $L (a,i) = (a,Li)$ with $ Li\not=i\,. $

\noindent
In case A a fixed point resolution
        is necessary
for the primary field labelled by $(a,i)$, whereas in case B a field
identification takes place.
Now condition [VIII] implies immediately that any fixed point of a
simple current of prime order must be resolved into $p$ fields with
multiplicity $m_{\alpha}=1$; hence in case A the field
$(a,i)$ is resolved into $p$ new primary fields $(a,i,\alpha)$,
$\alpha =1,2,...\,,p$. The field identification in case B
combines $p$ primaries $(a,i)$ (with fixed $a$, but distinct values
of $i$) into a single primary field of the \G-extended theory. In other
words, the $M_a^\H$ fields $(a,i)$ with given $a$
are combined into $M_a^{\H}/p$ new fields $(a,\ip)$,
where $\ip$ ($\ell=1,2,...\,,M_a^\H/p$) denotes some definite choice
among the labels $i$, reducing the label set by a factor $p$.
It follows that if all extensions were as in case A, the
multiplicities $m$ would always be equal to 1, and the total
number of fields would be equal to $|\S_a|$, the order of the
stabilizer of $a$. In contrast, case B amounts to a reduction of the
number of primary fields by a factor $p^2$, which is accompanied by
an enlargement of the multiplicity $m$ by a factor of $p$ due to the
sum over the $L$-orbit. (Note that $L$ must generate an orbit
of order $p$ on the resolved $\H$-fixed points, and under condition
[VIII] this is only possible if $|\H|$ contains a factor $p$.)
As a consequence, the number of primaries into which a fixed point
$a$ of the extension by $\G$ is resolved can in general be any integer
$|\S_a|/N^2$ with $N\in\Zbf$. Inspecting successive extensions allows
us to decide which of these possibilities is realized.

Let us now first consider case A.
For the Fourier basis we may take, without any loss of generality,
a subgroup $\M^{\H}_a \subset \M^{\G}_a$ with $|\M^\H_a|= |\M^\G_a|/p$.
Then any $g\in\M^\G_a$ can be written in the factorized form
$g=hw^\el$, where $h\in\M^\H_a$, and where $w$ is a coset
representative of a non-trivial element of the coset $\M^\G_a/\M^\H_a$
(we take the identity $\one\in\M^\G_a$ as the representative
for the trivial element of $\M^\G_a/\M^\H_a$); 
thus $w^\el\not\in\M^\H_a$ for $0<\el<p$. 

The $\M^\G_a$-characters then act as
  $$ \Psi^a_{(i,\alpha)}(hw^\el) =
  \Psi^a_{(i,\alpha)}(h)\Psi^a_{(i,\alpha)}
  (w^\el) =\Psi^a_{i}(h) \Psi^a_{(i,\alpha)}(w^\el)\,.$$
Here in the last equality we have used the fact that for
$h\in\M_a^\H$ the characters satisfy $\Psi_{(i,\alpha)}(h)=
\Psi_i(h)$, where the latter are characters of $\M_a^\H$.
{\small{
If furthermore $w^p =\one$, then we have
  $$ \M^\G_a = \M^\H_a \times \Zbf_p \,, $$
and the $\M^\G_a$-characters factorize as $\Psi^a_{(i,\alpha)}
(hw^\el)=\Psi^a_{i}(h)\Psi^a_{\alpha}(\el)$ into $\M_a^\H$-characters
$\Psi^a_i$ and $\Zbf_p$-characters $\Psi^a_{\alpha}$.}}

We can now write down formulas for the first extension, the
second extension, and the full extension.
The non-fixed orbits can be taken into account by restricting
the label $\alpha$ to a single value.
 To
make the notation unambiguous, resolved matrices will now
be distinguished by a superscript $\G$ and $\H$ (instead of
a tilde), and other quantities are labelled in the same way.
For the first extension we have
  $$ S^{\H}_{(a,i),(b,j)}={1\over \sqrt{M^{\H}_a M^{\H}_b}}
  \sum_{g \in \M^{\H}_a}\sum_{h \in \M^{\H}_b} S^{g,h\HH}_{a,b}\,
  \Psi^a_i(g)\, \Psi^b_j(h)^* \,, \eqn\EFirst $$
and for the second extension
  $$ S^{\G}_{(a,i,\alpha),(b,j,\beta)}=
  \sqrt{{M_a^\H M_b^\H \over M_a^\G M_b^\G}}
  \sum_{0 \leq n < p}\sum_{0\leq m<p} S^{n,m}_{(a,i),(b,j)}
  \Psi^a_{\alpha}(n)\,\Psi^b_{\beta}(m)^* \,. \eqn\ESecond $$
Note that the second extension requires a $\Zbf_p$-Fourier transform
even if $\M^{\G}$ is not the direct product of $\M^{\H}$ by $\Zbf_p$.

For the full extension we find
  $$ S^{\G}_{(a,i,\alpha),(b,j,\beta)}={1\over \sqrt{M^{\G}_a
  M^{\G}_b}}
  \sum_{g \in \M^\G_a}\sum_{h \in\M^\G_b} S^{g,h\GG}_{a,b}\,
  \Psi^a_{(i,\alpha)}(g)\, \Psi^b_{(j,\beta)}(h)^*\,. \eqn\EThird $$
Furthermore we have the relations (\cf\ \Soneone)
  $$ S^{\one,\one\HH}_{a,b}={|\H|\over
  \sqrt{|\S^{\H}_a|\,|\S^{\H}_b|}}\, S_{a,b}\eqn\EAA$$
and
  $$ S^{\one,\one\GG}_{a,b}={|\G|\over
\sqrt{|\S^{\G}_a|\,|\S^{\G}_b|}}\,
  S_{a,b} \,,  \qquad  
  S^{\one,\one}_{(a,i),(b,j)}={|\G|\over |\H|}\,
  \sqrt{|\S^{\H}_a|\,|\S^{\H}_b| \over |\S^{\G}_a|\,|\S^{\G}_b| }\,
  S^{\H}_{(a,i),(b,j)} \,. \eqn\ECC$$
Using these identities it is not difficult to show that
  $$ S^{g,h\GG}_{a,b} = {|\G|\over |\H|}\,
  \sqrt{|\S^{\H}_a|\,|\S^{\H}_b| \over |\S^{\G}_a|\,|\S^{\G}_b| }\,
  S^{g,h\HH}_{a,b}   \eqn\GHrel $$
when  $g\in \M^{\H}_a$ and $h\in \M^{\H}_b$.
Furthermore, if $g\in \M^{\H}_a$ but $h\not\in \M^{\H}_b$
(or vice versa), then $S^{g,h\HH}_{a,b}$ must vanish.

{\small{The details are as follows.
We substitute the last relation in \ESecond, and then we use \EAA:
  $$\eqalign{ S^\G_{(a,i,\alpha)(b,j,\beta)}&=
  {1\over \sqrt{M_a^{\G} M_b^{\G}}} {|\G|\over |\H|}\,
  {\sqrt{|\S^{\H}_a||\S^{\H}_b|}\over
\sqrt{|\S^{\G}_a||\S^{\G}_b|}}\,
  \sum_{g \in \M^{\H}_a}\sum_{h \in \M^{\H}_b} S^{g,h}_{a,b}[\H]\,
  \Psi^a_i(g) \Psi^b_j(h)^* \Psi^a_{\alpha}(0)
\Psi^b_{\beta}(0)^*\cr
  &+ {\sqrt{M_a^{\H} M_b^{\H}}\over \sqrt{M_a^{\G} M_b^{\G}}}
  \sum_{1 \leq n < p}\sum_{1 \leq m<p} S^{n,m}_{(a,i),(b,j)}
  \Psi^a_{\alpha}(n)  \Psi^b_{\beta}(m)^* \,. \cr} $$
(The first line is the 00-term.)
Note that the terms $S^{0,n}$ and $S^{n,0}$ with $ n \not= 0$ vanish.
To compare with \EThird\ we perform an inverse Fourier
transformation. Let us compute $ S^{g,h}_{a,b}[\G]$ for $h \in
\M^{\H}_a$. Suppose $h \in \M^{\H}_a$. Then
  $$ \eqalign{ S^{g,h}_{a,b}[\G]
  &={1\over \sqrt{M^{\G}_a M^{\G}_b}}
    \sum_{i,\alpha} \sum_{j,\beta} \Psi^a_{(i,\alpha)}(g)^*
    \Psi^b_{(j,\beta)}(h) S^\G_{(a,i,\alpha),(b,j,\beta)}\cr
  &={1\over \sqrt{M^{\G}_a M^{\G}_b}}
    \sum_{i,\alpha} \sum_{j,\beta} \Psi^a_{(i,\alpha)}(g)^*
\Psi^b_{j}(h)
    \bigg[ {1\over \sqrt{M_a^{\G} M_b^{\G}}}\, {|\G|\over |\H|}\,
    {\sqrt{|\S^{\H}_a||\S^{\H}_b|}\over
\sqrt{|\S^{\G}_a||\S^{\G}_b|}}\cr
  &~~\times \sum_{h'\in \M^{\H}_a}\sum_{h'' \in \M^{\H}_b}
    S^{h',h''}_{a,b}[\H]
    \, \Psi^a_i(h') \Psi^b_j(h'')^* \Psi^a_{\alpha}(0)
    \Psi^b_{\beta}(0)^*\cr
  &~~~~~~+ {\sqrt{M_a^{\H} M_b^{\H}}\over \sqrt{M_a^{\G} M_b^{\G}}}
    \sum_{1\leq n < p}\sum_{1\leq m<p} S^{n,m}_{(a,i),(b,j)}\,
    \Psi^a_{\alpha}(n)  \Psi^b_{\beta}(m)^* \bigg] \,.\cr}$$
Then the last term is projected out by the sum over $\beta$.
Furthermore $\Psi^a_{\alpha}(0)=\Psi^b_{\beta}(0)=1$.
Since $h'\in \M_a^{\H}$ we have $\Psi^a_i(h')=\Psi^a_{i,\alpha}(h')$.
The sum over $(j,\beta)$ yields $M_b^{\G} \delta_{h,h''}$ and the sum
over $(i,\alpha)$ $M_a^{\G} \delta_{g,h'}$.}}

In case B we have $\M^{\H}_a \supset
 \M^{\G}_a$,
and we can work with group characters $\Psi^a_{(i)}$ acting on
$a\in \M^{\H}_a$, where the labelling is such that the subset
$ \Psi^a_{(i_p)} $ forms a set of characters of $\M^{\G}_a$. Again we will
    include
the limiting case $\M^\G = \M^\H$, to
    allow also for
fields that are not fixed points. Of course, the formula for the first
extension remains the same as in case A. For the full extension we have
  $$ S^{\G}_{(a,i_p),(b,j_p)}={1\over \sqrt{M^{\G}_a M^{\G}_b}}
  \sum_{g \in \M^{\G}_a}\sum_{h \in \M^{\G}_b} S^{g,h\GG}_{a,b}\,
  \Psi^a_{i_p}(g)\,  \Psi^b_{j_p}(h)^* \,, \eqn\FullExt$$
and for the second extension
  $$ S^{\G}_{(a,i_p),(b,j_p)}={|\G|\over |\H|}\,
  {{|\S^{\H}_a||\S^{\H}_b|}\over {|\S^{\G}_a||\S^{\G}_b|}}\,
  S^{\H}_{(a,i_p),(b,j_p)} \,. \eqn\SecondExt $$
This is just \GenS\ with $m_i=1$ and $\Gamma=0$, since in the
second step there are no fixed points to resolve.
Note that all matrix elements of $S^{\H}_{(a,i_p),(b,j_p)}$ are the
same on the $L$-orbits, so that the answer does not depend on which
element we select for $i_p$.

Combining this information, we obtain again
the result \GHrel, except that this time it completely
determines $S^{g,h\GG}$, i.e.\ is valid for all $g\in\M^{\H}_a$ and
all $h\in\M^{\H}_b$.
{\small{The calculation goes as follows.
Substituting the first extension \EFirst\ into \SecondExt\ we get
  $$ S^{\G}_{(a,i_p),(b,j_p)}={|\G|\over |\H|}\,
  {\sqrt{|\S^{\H}_a||\S^{\H}_b|}\over
\sqrt{|\S^{\G}_a||\S^{\G}_b|}}\,
  {1\over \sqrt{M^{\H}_a M^{\H}_b}}
  \sum_{g \in \M^{\H}_a}\sum_{h \in \M^{\H}_b} S^{g,h}_{a,b}[\H]\,
  \Psi^a_{i_p}(g) \Psi^b_{j_p}(h)^*  \,. $$
Comparing with \FullExt\ shows that $g,h\in\M^{\H}_a\setminus
\M^{\G}_a$ do not contribute, and hence (using \smmm)
  $$ \eqalign{ S^{g,h}_{a,b}[\G]&=
  {\sqrt{M^{\G}_a M^{\G}_b}\over \sqrt{M^{\H}_a M^{\H}_b}} {|\G|\over
  |\H|}\, { {|\S^{\H}_a||\S^{\H}_b|}\over  {|\S^{\G}_a||\S^{\G}_b|}}\,
  S^{g,h}_{a,b}[\H]\cr
  &= {\sqrt{M^{\G}_a M^{\G}_b}\over \sqrt{M^{\H}_a M^{\H}_b}}
  {|\G|\over |\H|}\, {m^{\H}_a m^{\H}_b\, \sqrt{M^{\H}_aM^{\H}_b}
  \over m^{\G}_a m^{\G}_b \sqrt{M^{\G}_aM^{\G}_b}}\,
  {\sqrt{|\S^{\H}_a||\S^{\H}_b|}\over
  \sqrt{|\S^{\G}_a||\S^{\G}_b|}}\, S^{g,h}_{a,b}[\H]
  = {|\G|\over |\H|}\, {\sqrt{|\S^{\H}_a||\S^{\H}_b|}\over
 \sqrt{|\S^{\G}_a|| \S^{\G}_b|}}\, S^{g,h}_{a,b}[\H]\,, \cr} $$
where in the last step we used $m_a^{\G}=m_a^{\H}$, since
none of the multiplicities can change in a homogeneous extension of
prime order.}}

In principle it might happen that cases A and B occur
simultaneously for a given extension. Then there are matrix elements
of $S^{\G}$ between fields $a$ and $b$ with
$\M^{\G}_a \subset \M^{\H}_a$ and $\M^{\G}_b \supset \M^{\H}_b$. The
analysis of this mixed case is essentially the same as in the
previous case, and the result is once again \GHrel, with the
restriction that $S^{g,h}=0$ whenever $h\not\in \M^{\H}_b$.

{\small{For the second extension we get in the mixed case
  $$ S_{(a,i_p),(b,j,\beta)}^{\G}={|\G|\over |\H|}\,
  {{|\S^{\H}_a||\S^{\H}_b|}\over {|\S^{\G}_a||\S^{\G}_b|}}\,
  S^{\H}_{(a,i_p),(b,j)} \,. $$
This is just \GenS\ between a non-fixed point and a fixed point. The
correction matrix $\Gamma$ appears only if both $a$ and $b$ are
fixed points of the second extension. For the full extension we find:
  $$ S_{(a,i_p)(b,j,\beta)}^{\G}={1\over \sqrt{M^{\G}_a M^{\G}_b}}
  \sum_{g \in \M^{\G}_a}\sum_{h \in \M^{\G}_b} S^{g,h}_{a,b}[\G]\,
  \Psi^a_{i_p}(g)  \Psi^b_{(j,\beta)}(h)^* \,. $$
The rest of the computation closely resembles the previous case.}}

All these cases are summarized by the formula
  $$ S^{\G}_{(a,I),(b,I')}={1\over \sqrt{M^{\G}_a M^{\G}_b}}
  \sum_{g \in \M^{\G}_a}\sum_{h \in \M^{\G}_b} S^{g,h\GG}_{a,b}\,
  \Psi^a_{I}(g)\,\Psi^b_{I'}(h)^*  \eqn\FullExtGen $$
for the full extension,
where $I$ now stands for either the combination of labels
$(i,\alpha)$ or the single label $i_p$, or just the label $i$ or
$\alpha$, and  analogously for
$I'$. Furthermore we always have the relation \GHrel. In case A this
only gives us part of $S^{g,h\GG}$ whereas for case B it gives us all of
$S^{g,h\GG}$. Note that even though \FullExtGen\ is universal, the
factor $1/\sqrt{M^{\G}_a}$ for case B is by a factor $p$ larger than
in case A.

\section{Condition [VII]: Commutativity of extensions}

\noindent
In the previous section we have analyzed the consequences of
condition [VII] by comparing two successive extensions to the full
extension.
Another aspect of condition [VII] is that two successive extensions
should commute whenever each of them can be performed as the
first extension. To check this commutativity,
we have to compare the embedding chains
  $$ \G = \H_1 \times \H_2 \supset\H_1 \supset \{\one\}\qquad
\hbox{~and~}
  \qquad \G = \H_1 \times \H_2 \supset\H_2 \supset \{\one\} \,. $$

Consider two fields $a$ and $b$ with stabilizer $\S=\S_1 \times
\S_2$, all with implicit labels $a$ respectively $b$. For simplicity
we will assume that case A applies in all cases, so that $|\M|=|\S|$.
Without loss of generality we may then choose for the group $\M_i$
the stabilizer $\S_i$. Requiring that each of the two embedding
chains yields the same answer leads to the condition
  $$ S^{g,h\GG}={|\G|\over |\H_i|}\,
  \sqrt{{|\S^{\H}_i||\S^{\H}_i|}\over {|\S^{\G}||\S^{\G}|}}\,
  S^{g,h\HH_i} \eqn\GroupDiagonality $$
when either $g$ or $h$ are restricted to the subgroup $\M_i$.
In particular $S^{g,h\GG}$ vanishes when $g$ and $h$ are from
different factors of $\G$. This holds equally well for any other
decomposition of $\G$.

\section{The matrices $S^J$}

At this point it is worth stressing that so far there was no need to
specify the
      discrete
abelian group $\M_a$ (except for the restriction that the
groups associated to successive embeddings are contained in each other).
Rather, choosing a particular group $\M_a$ is merely a matter of
convenience. However, while for any fixed point resolution the Fourier
 transformation \Fourier\
can be performed for any arbitrary choice of
$\M_a$, this manipulation is not likely to lead to useful results unless
a clever choice of $\M_a$ is made.

Now the results of the previous sections inspire us to make
the following {\it ansatz} for $S^{g,h}_{a,b}$.
First of all, we identify the elements $g$ and $h$ of the groups
$\M_a$ and $\M_b$ with elements $J$ and $K$ of the
stabilizer $\S_a$ or $\S_b$, respectively. This is obviously possible
if all multiplicities $m_{a,i}$ are equal to 1, since in that case
$M_a=|\S_a|$. Otherwise these numbers differ by a square
(because of condition [VIII]), and one can always find a subgroup of
$\S_a$ that
has the right size. This subgroup may not be unique, but we will soon
make a canonical choice. Now we
      make the ansatz
  $$S^{J,K\GG}_{a,b} :={|\G| \over\sqrt{ |\S_a|\,|\S_b|}
  }\,\delta_{JK}\, S^J_{a,b} \,.  \eqn\Sdef $$
This defines a matrix
 $S^J_{a,b}$
for each current, which is however
independent of the extension one considers. The precise definition
of $S^J$ can  already be obtained by considering the minimal
extension for which it appears, namely
the extension by $J$ itself (or more precisely, the discrete group
$\H_J$ it generates). In that case \Sdef\ reads
  $$S^{J,K\HHJ}_{a,b} =\delta_{J,K}\,S^J_{a,b} \,, $$
since $\G=\H_J$, and  the identification of $J$ with an element of
$\S_a$ and one of $\S_b$ means that $S^J_{ab}=0$ unless $Ja=a$ and
$Jb=b$.  Thus $J \in\S_a\cap \S_b$, which implies that
$\H_J=\S_a=\S_b$\rlap.\foot{In principle it could happen that $J \in
\S$, but $J \not \in \M$. Since the Fourier transforms are
defined using  $\M$, this would imply that $S^J_{a,b}$ is then
not defined for all primary fields $a$. It can be shown (see the appendix)
that this situation -- which can only be due to successive resolutions
according to case B above -- can never arise within a single cyclic
group. (Also, one can argue that these matrix elements
are anyway never needed in further extensions.) Hence
if we extend the algebra only by $J$ (and its powers), then all fixed
points are fully resolved into
         fields that can be described in terms of $\S_a$.}

Note that for $\G = (\Zbf_2)^n$ this structure follows completely
from the foregoing discussion. For each single $\H_J=\Zbf_2$
extension by a current $J$ we have
  $$ S^{0,J}=0=S^{J,0}\eqn\AssThree$$
and
  $$ S^{0,0\HHJ}_{a,b}= {|\H_J| \over
  \sqrt{|\S^{\H_J}_a|\,|\S^{\H_J}_b|}}\,
  S_{a,b} \,, \qquad  
  S^{J,J\HHJ}_{a,b}= {|\H_J| \over
  \sqrt{|\S^{\H_J}_a|\,|\S^{\H_J}_b|} }\,
  S^J_{a,b}\ (=S^J_{a,b}) \,. \eqn\AssTwo$$
The last of these equalities is the definition of $S^J$, while the
others follow from  \Soneone\ and tell us that $S^\one_{a,b}=S_{a,b}$.
Using \GroupDiagonality, this implies immediately that \Sdef\ holds
for any further extension. Therefore the non-trivial assumption in
\Sdef\ is that the matrices $S^{J,K}$ vanish for distinct currents $J$ and
$K$ even if they belong to the same cyclic subgroup of \G\ and have
the same order.

Now we substitute the ansatz \Sdef\ in conditions [II]--[V], derived
for the most general form of the resolution procedure. We consider
first the defining matrices obtained for the minimal extension. We
find:\smallskip

\leftline{$\underline{\hbox{Condition [II]:}}$}
  $$ S_{a,b}^J=S_{b,a}^{J^{-1}} \,. \eqn\PartSym$$

\leftline{$\underline{\hbox{Condition [III]:}}$}
 $$ \sum_b S_{a,b}^J\, (S_{b,c}^{J^{-1}})^* = \delta_{bc}\,.  $$
When combined with \PartSym, this implies that $S^J$ is unitary.
\smallskip

\leftline{$\underline{\hbox{Condition [IV]:}}$}
We obtain $\sum_b S_{a,b}^J S_{b,c}^J \delta_{JK}^{}
= \hat C_{a,c}^{} C^c_{J,K}$, where $C^c_{J,K}$ is a unitary matrix.
The condition [IV] says that it must also be diagonal,
so that its diagonal matrix elements must be phases. Hence we get
  $$ \sum_b S_{a,b}^J S_{b,c}^J = \hat C_{a,c}\,\eta^J_c $$
with some phases $\eta^J_c$.

The fact that $C^c_{J,K}$ is diagonal implies that
${1\over M_a} \sum_{i,j} \Psi^a_{i}(J)C^c_{i,j}(J)\Psi^a_{j}(K)^* =
\eta_c^J \delta_{JK}$. 
Multiplying with $\Psi^a_k(K)$ and summing over $K$ then gives
  $$ \sum_i \Psi^a_{i}(J)\,C^c_{i,k} = \eta_c^J\, \Psi^a_k(J) \,. $$
In the sum on the left hand side only a single term survives. Define
$k^c$ by
  $$ C^c_{j,k}= \delta_{j,k^c} \,.$$
Then we get
  $$\Psi^a_{k^c}(J)=\eta_c^J\, \Psi^a_k(J) \,. $$
This allows us to write $\eta^J_c$ (considered as a function of $J$)
as a ratio of two group characters.
It then follows that it enjoys the group property
  $$ \eta_c^{J_1}\eta_c^{J_2}=\eta_c^{J_1J_2} \,. $$
In particular $(\eta_c^J)^N = 1$ if $J$ has order $N$. This implies
in  particular that $\eta_c^J=-1$
is allowed only for currents of even order that are not
themselves a square of other currents.

The property $C^c_{i,j}=C^{c^*}_{j,i}$ implies that
$\eta^J_c=(\eta^J_{c^*})^*$.

\smallskip
\leftline{$\underline{\hbox{Condition [V]:}}$}
  $$ \sum_b S_{a,b}^J T_{b,b}^{} S_{b,c}^J \delta_{JK}=
  T^{-1}_{a,a} S^J_{a,b} T^{-1}_{b,b} \ .$$
We thus find that $S^J$ must form a unitary representation of
the modular group on the fixed points of $J$. This means that
$S^J(S^J)^{\dagger}=\unit$, $(S^JT)^3=(S^J)^2$ and,
      due to $\eta^J_c=(\eta^J_{c^*})^*$,
$(S^J)^4=\unit$;
note,  however, that it is {\it not} required that $S^J$ must
be symmetric, nor does $(S^J)^2$ have to be a permutation.
\smallskip
\leftline{$\underline{\hbox{Condition [VI]:}}$}
{}From the general result \eSFS\ we get directly
  $$ \ee^{2\pi \ii Q_L(b) } S^J_{a,b}\, \delta_{JK}=
  \sum_M {\cal F}_{\!J,M}^a(L)\,S^M_{La,b}\, \delta_{MK} \,. $$
This implies that the matrix ${\cal F}_{\!J,M}$ must be diagonal.
Unitarity of $S^J$ then requires that the elements of $\cal F$
should be phases. Hence
  $$ {1\over M_a} \sum_i \Psi^a_{i}(J)\,\Psi^a_{Li}(M)^* =F(a,L,J)\,
  \delta_{JM}  \eqn\PhaseDef$$
with certain phases $F(a,J,K)$. The relation for $S^J$ that we get is
$$S^J_{La,b} = F(a,L,J)\, \ee^{2\pi \ii Q_L(b) }\, S^J_{a,b}
\,.\eqn\PHases$$
Multiplying \PhaseDef\ with $\Psi^a_k(M)$ and summing over $M$ we
then get $F(a,L,J)^* \Psi^a_k(J) = \Psi^a_{L^{-1}k}(J) $, \ie
  $$  F(a,L,J)^*  = \Psi^a_{L^{-1}k}(J) \,/\, \Psi^a_k(J) \,. $$
Since this is a ratio of group characters, the phase satisfies
the group property
  $$F(a,L,J_1)\,F(a,L,J_2)= F(a,L,J_1J_2) \,.\eqn\GroupProp $$

The quantities $F(a,K,J)$ decide whether the current $K$ acts
non-trivially
on the labels of the resolved fixed points of $J$. This action
is trivial if and only if $F(a,K,J)=1$, but the value of $F$
is relevant only if $a$ is fixed by $K$. Indeed, in the appendix
we show that the value of $F$ on other fields can be modified by
conjugating $S^J$ by a diagonal unitary matrix. This changes both
$F$ and $\eta$, and shows that only the value of these parameters on
fixed points and self-conjugate fields, respectively, is relevant
information. In the appendix these transformations are used to
choose a convenient `gauge' such that $\eta=1$ on fields that are
not self-conjugate and that both $\eta$ and $F$ are constant on
simple current orbits, and such that $F$ satisfies the group
property also with respect to its second argument.

Furthermore, as is shown in the appendix, with this choice the phases
$F$ for integral spin currents $J$ and $K$ are equal to
  $$  F(a,K^pJ^q,K^nJ^m)=F(a,K,J)^{pm-nq} \,,\eqn\Fsymm $$
so that on each $\Zbf_N \times \Zbf_M$ subgroup of the center,
$F$ can be completely expressed in terms of a single phase.

We are now in a position to determine precisely under which
conditions case B of section 4.8 applies. Clearly this
situation occurs whenever two currents $J$ and $K$ from different
orbits both fix a field $a$, with $F(a,K,J) \not =1$. The
relation \Fsymm\ guarantees that the latter property is symmetric
in $J$ and $K$, so that it does not matter whether we first extend
by $J$ and then by $K$ or the other way round.
Inspired by these observations we define for each field
the {\it untwisted stabilizer} $\U_a$ as
  $$ \U_a:= \{ J \in \S_a \,|\, F(a,K,J)=1 \ \hbox{for all ~} K
  \in \S_a\} \,.\eqn\UntwistedStabilizer $$
Because of the group property \GroupProp\ this is a subgroup of
$\S_a$. Because the defining relation
is symmetric in $J$ and $K$,   $|\S_a/\U_a| $ is always a square.
The discussion leading to \FullExtGen\ shows that the
multiplicity factor(s) $m_a$ should be chosen equal to
the integer  $\sqrt{|\S_a/\U_a|}$, and the number of fields
should be reduced by the square of this factor, so that it is
precisely $|\U_a|$. Consequently the untwisted subgroup $\U_a$ rather
than the full stabilizer $\S_a$
is the natural group to use for the Fourier decomposition.

A few other observations which show that the untwisted stabilizer is
a rather natural concept are the following. First, as follows from
the result \SJFSJ\ below, not only $J\in\S_b$, but even $J\in\U_b$ is
a necessary (though not sufficient) condition for $S^J_{a,b}$ to be
non-vanishing. Second, as we will show in the appendix, the condition
$F(a,K,J) = 1$ is symmetric in $J$ and $K$, and
the same group $\U_a$ is obtained if one replaces the $F$ in
\UntwistedStabilizer\ by $\bar F$ (which is the analogue of $F$
for the action of the current on the second
index of $S^J$). Finally, the groups $\U_a$ are invariant under the
transformations \PhaseRot\ which respect the conditions on the
matrices $S^J$.

We have now gathered all the ingredients for the
formula for the matrix $\tilde S$. What is still missing
is a proof that the matrices $S^{J,K}$ defined in \Sdef\ satisfy all
the relevant requirements also for  extensions that are not
minimal. Since these matrices are expressed in terms of the
matrices $S^J$ for the minimal extension, this should now
follow from the conditions on $S^J$. Before examining this,
we summarize all these conditions and write down
an explicit the formula for $\tilde S$.

\chapter{The formula for $\tilde S$}

\section{The properties of the matrices $S^J$}

\noindent
The observations in the previous section can be summarized as follows.
We are given a conformal field theory, with a set of mutually
local integer spin simple currents forming a discrete group $\G$. To
resolve the fixed points in the extension by $\G$ we need at least
the following data.

\noindent
For every simple current $J$  we are given a matrix $S^{J}$ that
satisfies

\item\bullet $\{1\}$ \ $S^{J}_{a,b} = 0\,$ if $\,Ja\not=a$ or
$Jb\not= b$.

\noindent
{}From now on $S^J$ refers only to
     the
restriction to the fixed points.
Thus $S^J$ is a non-trivial matrix only if $J$ has fixed points,
which can happen only if $J$ has integer or half-integer spin.

\item\bullet $\{2\}$ \ $S^{J}$ is unitary.

\item\bullet $\{3\}$ \ $S^{J}$ satisfies $(S^J T^J)^3 = (S^J)^2 $.

\noindent
Here $T^J$ is the $T$-matrix of the
     unextended
theory, restricted to the fixed
points of $J$.

\noindent
We require the following simple current transformation rules:

\item\bullet $\{4\}$ \
   $S^{J}_{Ka,b} = F(a,K,J)\,\ee^{2\pi\ii Q_K(b)}\,S^{J}_{a,b}\,,$

   $~~~~~~~~~S^{J}_{b,Ka}=\bar F(a,K,J)\,\ee^{2\pi\ii Q_K(b)}\,
   S^{J}_{b,a}\,$.

\noindent
Here $K$ is any other simple current that is local
with respect to $J$, and $Q_K(b)$ is
the monodromy charge of $b$ with respect to $K$, defined for the
unextended theory the matrix $S\equiv S^\one$ as in \ChargeDefinition.
This is not merely a definition of $F$ and $\bar F$, but implies that
they do not depend on $b$. In addition we require

\item\bullet $\{4a\}$ \ $F(a,K,J_1)\, F(a,K,J_2)=F(a,K,J_1J_2)$

\noindent
for all currents in $\S_a$.
Owing to the group property $\{4a\}$ we have $F(a,K,\one)=1$
(this also follows directly from condition $\{4\}$ because
for $J=\one$, $\{4\}$ is the usual simple current relation
\CSR\ for the matrix $S^\one\equiv S$).

Using the functions $F(a,K,J)$ we define for each primary field
$a$ the untwisted stabilizer $\U_a$ as in
\UntwistedStabilizer.

\noindent
The charge conjugation conditions can be stated in terms of
a matrix $\eta$ defined by
  $$(S^J)^2 =\eta^J C^J \ ,$$
where $C^J$ is the charge conjugation $C$ of the
     unextended
theory
restricted to the fixed points of $J$. Note that if $a$ is a fixed
point of $J$, then so is its conjugate (denoted $a^*$ in the
following), so that the restriction makes sense.
The matrices $\eta$ must satisfy the following conditions:

\item\bullet $\{5a\}$ \ $\eta^J \hbox{~is diagonal}$.

\item\bullet $\{5b\}$ \
$\eta^{J_1}_{a,a}\eta^{J_2}_{a,a}=\eta^{J_1J_2}_{a,a}
\ \  \ \hbox{for all~~} J_i \in \U_a$.

\item\bullet $\{5c\}$ \ $\eta^J C^J=C^J (\eta^J)^{\dagger}\,$.

\noindent
Thus the numbers $\eta^J_{a,a}$ are phases,
conjugate fields have complex conjugate $\eta$-values, and
on self-conjugate fields $\eta$ can take only the values $\pm1$. Note
that condition $\{5c\}$ is equivalent to $(S^J)^4=\unit$.
Condition $\{5b\}$ is required only on the untwisted
stabilizer $\U_a$ of $a$, not on
the full stabilizer. Note also that $\eta^\one_{a,a}=1$.

\noindent
Furthermore we demand that $S^J$ should satisfy

\item\bullet $\{6\}$ \ $S^{J}_{a,b} = S^{J^{-1}}_{b,a}\,.$

Above we have postulated a lot of structure that should
be present in any rational conformal field theory. One may
wonder whether this structure is something completely new
or whether it is already available in the usual data that
 are associated
to a rational conformal field theory,
e.g.\ those which appear in the polynomial equations \Mooresei.
Indeed, we conjecture that the matrix $S^J$ coincides with
the matrix that describes the modular transformation
properties of the one-point function on the torus with insertion
of the simple current $J(z)$. A proof of this conjecture is
however beyond the scope of this paper.
In section 6 we will show that for the important special case of
WZW-models natural solutions to all these conditions can be written down
explicitly.

\section{The main formula}

We work here with the group characters
of the untwisted stabilizer, which have the usual properties, see
section 4.1.

The primary fields of the extended theory can be described as follows.
Each fixed point $a$ of the unextended theory is resolved into
$|\U_a|$ distinct fields, which are labelled by the group characters
of the untwisted stabilizer $\U_a$. Then the following is the formula for
the modular matrix $\tilde S$:
  $$\tilde S_{(a,i),(b,j)}=
  {|\G| \over \sqrt{|\U_a|\,|\S_a|\,|\U_b|\,|\S_b|}}\,
  \sum_{J \in \G} \Psi^a_i(J)\, S^J_{a,b}\, \Psi^b_j(J)^*\,.  \eqn\XX$$
Here the  summation is formally over all $\G$, but in fact the
only contributing terms are those
with $J \in \U_a \cap \U_b$. In particular, if a primary field $a$ is not a
fixed point of any current, then $\U_a=\{\one\}$, and only $S^\one$
(the modular matrix $S$ of the unextended theory) contributes.

The formula \XX\ follows directly from the Fourier decomposition
\Fourier\ with $\M_a=\U_a$ and the diagonality assumption \Sdef,
which in its turn is strongly suggested by the arguments in
sections 4.8 and 4.9.

\section{Phase rotations}

As mentioned in the previous section,
all conditions on $S^J$ are respected by the `gauge' transformation
  $$ S^J \mapsto D^J S^J (D^J)^{\dagger} \,, \eqn\PhaseRot $$
where $D^J$ is a diagonal unitary matrix which,
in order to preserve  $\{6\}$, satisfies
  $$D^J=(D^{J^{-1}})^*\,. \eqn\DJinv $$
A sufficient condition for preserving the group properties
of $\eta$ and $F$, $\{5b\}$ and $\{4a\}$ is
  $$ D^{J_1} D^{J_2}= D^{J_1J_2} \,. \eqn\Dgrp$$
However, $F$ and $\eta$ change only
by ratios of the matrix elements of $D^J$, and therefore
these latter conditions are necessary only for those ratios.
These ratios are between conjugate fields or fields on the
same simple current orbits. There are thus many
phase rotations that are not restricted by \Dgrp.

This implies that in any case for the description of fixed point
resolution we will have to deal with ambiguities.
To see the effect of these phases, let us
absorb them into the group characters by defining
  $$\tilde \Psi^a_i(J):= D^J_{a,a}\, \Psi^a_i(J) $$
for each $a$. The new functions
    (which are generically not group characters any more)
can be expanded with respect to the old ones,
  $$  \tilde \Psi^a_i(J)=\sum_j d^a_{ij} \Psi^a_j(J) \,. $$
To compute the coefficients $d^a_{ij}$ we multiply both sides with
$\Psi^a_k(J)^*$ and sum over $J$:
  $$ d^a_{ik} = {1\over |\U_a|} \sum_J D^J_{a,a}\,
  \Psi^a_i(J)\,\Psi^a_k(J)^* \,.  $$
Because of \DJinv\ and the inversion property of the characters,
the numbers $d_{ik}$ are real. Furthermore $d$ is unitary:
  $$\eqalign{ \sum_k d^a_{ik}(d^a_{lk})^*  &=
  {1\over |\U_a|^2}\sum_k\sum_{K,J} D^J_{a,a}
\Psi^a_i(J)\Psi^a_k(J)^*
  (D^K_{a,a})^* \Psi^a_l(K)^*\Psi^a_k(K)
  \cr&=  {1\over |\U_a|} \sum_J  D^J_{a,a} \Psi^a_i(J)\,
  (D^J_{a,a})^* \Psi^a_l(J)^* = \delta_{il}\,. \cr} $$
Therefore it is orthogonal.

The new matrix $\tilde S$ is related to the old one as
  $$ \tilde S_{[{\rm new}]}=d\, \tilde S_{[{\rm old}]}\, d^{\dagger}
  \,, $$
where $d$ is a block-diagonal matrix with blocks acting on each
fixed point space. Since $d$ is unitary, it preserves unitarity of
$\tilde S$. Since it is orthogonal, it preserves symmetry of $\tilde
 S$, as well
as $(\tilde S\tilde T)^3=\tilde S^2$ and $\tilde
S^4=\unit$.

The transformation \PhaseRot\
      thus
generates a rotation among the primary fields into which a fixed point is
resolved that respects all modular group representation properties.
These transformations change the fusion rules unless the same
transformation is applied to all fields simultaneously, in a sense
         that we will formulate
more precisely in section 5.5. Since $d$ is a generic
orthogonal matrix, not necessarily a permutation,
        in general
the transformation does not even respect
integrality of the fusion coefficients.
However, because of conditions $\{5b\}$ and $\{4a\}$ which must
be respected by $D^J$, it will leave two important aspects of the
fusion rules unaffected, namely the charge conjugation matrix
and simple current fusion rules. Although both can change, they
will in any basis have the correct form.

The existence of this
freedom implies that given a set of matrices $S^J$ satisfying all 6
conditions, one may discover that the matrix $\tilde S$ does not
yield correct fusion rules. This does not necessarily imply that the
fixed point resolution procedure has failed, but merely that
one may have to choose
a different basis using the phase rotations. In the application
to WZW-models these considerations will
not play a role, because there is a canonical basis for $S^J$
which -- by inspection -- yields correct fusion rules.

Given a basis choice that yields proper
fusion coefficients, there are still further phase rotations
one can make that do not affect the integrality of the fusion
rule coefficients. Namely, it is easy to show that
if the matrices $D^J$ satisfy \Dgrp, then
the matrix $d^a_{ij}$ is a permutation. If it
depends on $a$ it changes the fusion rules, but does not affect their
integrality.

\section{Proofs}

\noindent
Most of the properties of $\tilde S$ were already proved in the
previous section. In particular the presence of the terms $S^\one$
guarantees that condition [I] is satisfied, and $\{6\}$ in
combination with the group property of the characters under
inversion imply that $\tilde S$ is symmetric.

{\small{
Calculation:
  $$\eqalign{\tilde S_{(a,i),(b,j)} \ &=
  {|\G| \over \sqrt{|\U_a||\S_a||\U_b||\S_b|}}
    \sum_{J \in \G} \Psi^a_i(J) S^J_{a,b} \Psi^b_j(J)^*\cr
  [\hbox{using \Inv}]&= {|\G| \over \sqrt{|\U_a||\S_a||\U_b||\S_b|}}
    \sum_{J \in \G} \Psi^a_i(J^{-1})^* S^J_{a,b} \Psi^b_j(J^{-1})\cr
  [\hbox{using [6]}]&= {|\G| \over \sqrt{|\U_b||\S_b||\U_a||\S_a|}}
    \sum_{J \in \G}  \Psi^b_j(J^{-1})S^{J^{-1}}_{b,a}
\Psi^a_i(J^{-1})^* \cr
  [\hbox{re-ordering  $\G$-summation}]&= {|\G|\over
\sqrt{|\U_b||\S_b|
    |\U_a||\S_a|}} \sum_{J \in \G}  \Psi^b_j(J)S^J_{b,a}
\Psi^a_i(J)^*
   = \tilde S_{(b,j),(a,i)} \,. \cr }$$ }}

The most crucial and non-trivial check is unitarity. This  amounts to
proving that the matrices $S^{J,K}$ of \Sdef\ satisfy \SghUnit.
This can easily be done, but instead we prefer to give here a direct
proof. Define
  $$ N(a,b):={|\G| \over \sqrt{|\U_a|\,|\S_a|\,|\U_b|\,|\S_b|}} $$
It is straightforward to show that
  $$ \eqalign{\sum_{b,j}& \tilde S^{}_{(a,i),(b,j)} \tilde
  S_{(b,j),(c,k)}^* =\sumRQ b
  N(a,b)N(b,c)\, |\U_b|\!\!\! \sum_{J \in\U_a\cap \U_b\cap\U_c}
  \!\!\Psi^a_i(J)\,\Psi^c_k(J)^*\, S_{a,b}^J\, (S_{c,b}^{J})^* \,.
  \cr}$$
The symbol $\sumRQ{\!\!\!}$ indicates a sum over one
representative out of each $\G$-orbit with $Q_{\G}(b)=0$.

{\small{
Calculation:
  $$ \eqalign{\sum_{b,j}&\tilde S_{(a,i),(b,j)}\tilde
S_{(b,j),(c,k)}^* \cr
  [\hbox{Symmetry}]&=
    \sum_{b,j} \tilde S_{(a,i),(b,j)} \tilde S_{(c,k),(b,j)}^*\cr
  &=\sumRQ b N(a,b)N(b,c) \sum_J \sum_{J'}
    \sum_j\Psi^a_i(J) S_{a,b}^J \Psi^b_j(J)^* \Psi^c_k(J')^*
    (S_{c,b}^{J'})^* \Psi^b_j(J') \cr
  [\hbox{Using \Compl}] &=\sumRQ b N(a,b)N(b,c)\, |\U_b| \sum_J
    \sum_{J'} \Psi^a_i(J) S_{a,b}^J  \Psi^c_k(J')^* (S_{c,b}^{J'})^*
\delta
    ^{\U_b}_{JJ'} \cr
  &=\sumRQ b N(a,b)N(b,c)\, |\U_b| \sum_{J \in\U_b}
    \Psi^a_i(J)  \Psi^c_k(J)^* S_{a,b}^J (S_{c,b}^{J})^*  \,. \cr}$$
}}

Now we sum over a full orbit of $b$ rather than just a
representative.
Each term in such an orbit gives the same contribution
because, due to $\{4\}$,
  $$ S_{a,Kb}^J(S_{c,Kb}^J)^* = \ee^{2\pi \ii(Q_K(a)-Q_K(c))}
  F(b,K,J)F(b,K,J)^*\, S_{a,b}^J(S_{c,b}^J)^*  \eqn\bCancel $$
and $Q_{\G}(a)=Q_{\G}(c)=0$.
Hence we can extend the sum over the full orbit, and divide by the
orbit length $|\G|/|\S_b|$. After this operation we get a factor
$N(a,b)N(b,c) |\U_b||\S_b|/|\G|=N(a,c)$, and in particular
(except for the range of the $J$-summation) all dependence on $b$
disappears.
We wish to sum also over all $b$ with $Q_{\G}(b) \not = 0$, because
only in that case we can use the unitarity relation for $S^J$. To do
so,we observe that
   $ S_{a,b}^J= F(a,K,J)^*\, S_{Ka,b}^J$ if $Q_{K}(b) = 0 $.
Hence we may replace $S_{a,b}^J$ by
$ {1\over |\G|} \sum_{K \in \G}  F(a,K,J)^*S_{Ka,b}^J.$
Suppose we now consider a field $b$ with non-zero monodromy charge.
Then we have
  $$ {1\over |\G|} \sum_{K \in \G}  F(a,K,J)^*\, S_{Ka,b}^J =
  {1\over |\G|} \sum_{K \in \G} \ee^{2\pi \ii Q_K(b) }S_{a,b}^J = 0
  \,.$$
Clearly we are therefore allowed to extend the sum over $b$ to all
fields.
{\small{
Calculation: Using the convention that
  $$ \Psi^a_i(J) = 0 \hbox{~~if~~} J \not \in \U_a \,, \eqn\Vanish$$
we get
  $$ \eqalign{ \sum_b
  &N(a,b)N(b,c)\, |\U_b|{ |\S_b|\over|\G |^2} \sum_{K \in \G}
  \sum_{J\in\U_b}
    F(a,K,J)^* \Psi^a_i(J) \Psi^c_k(J)^* S_{Ka,b}^J (S_{c,b}^{J})^*
  \cr &= \sum_b {1\over \sqrt{|\U_a| |\U_c| |\S_a| |\S_c|} }
    \sum_{K \in \G} \sum_{J \in \U_b} F(a,K,J)^*
    \Psi^a_i(J)  \Psi^c_k(J)^* S_{Ka,b}^J (S_{c,b}^{J})^*\cr
  \hbox{Using \Vanish} &= \sum_b
    {1\over \sqrt{|\U_a| |\U_c| |\S_a| |\S_c|} }
    \sum_{K \in \G}\left\{ \sum_{J \in\U_a\cap\U_c} \! - \!\!
    \sum_{\scriptstyle J \in\U_a \cap\U_c \atop\scriptstyle J \not
    \in\U_b} \right\} F(a,K,J)^*
    \Psi^a_i(J)  \Psi^c_k(J)^* S_{Ka,b}^J (S_{c,b}^{J})^* \,.\cr} $$
    }}%
Now there is still one obstacle preventing us from performing
this summation over $b$, namely the restriction of $J$ to $\U_b$.
Note, however, that $S^{J}_{Ka,b}$ vanishes if $J\not\in \U_b$.
This is clearly true  for $J\not \in \S_b$ since then by $\{1\}$ we
have $S^J_{a,b}=S^J_{b,c}=0$. Consider now $J\in \S_b$, but
$J \not\in \U_b$. Then
there exists a current $L \in \S_b$ so that $F(b,L,J) \not = 1$.
Thus the fact that $b=Lb$ and $Q_L(Ka)=0$ imply that
  $$ S^J_{a,b} = S^J_{a,Lb} = F(b,L,J)\,\ee^{2\pi\ii Q_L(a)}
  S^J_{a,b}  = F(b,L,J)\, S^J_{a,b} \,, \eqn\SJFSJ $$
and hence $S^J_{a,b}$ has to vanish.
Hence we may extend the sum over $J$ to all of $\G$, but due to
the restriction of the sum to $\U_a \cap \U_b \cap \U_c$ we are
finally left with a sum over $\U_a  \cap \U_c$.
Now we can finally sum over $b$, using unitarity of $S^J$, to obtain
  $$ \sum_{b,j} \tilde S^{}_{(a,i),(b,j)} \tilde S_{(b,j),(c,k)}^*=
  {1\over \sqrt{|\U_a|\, |\U_c|\, |\S_a|\, |\S_c|} } \sum_{K \in \G}
  \sum_{J \in \U_a \cap \U_c} \!\!\!\!\! F(a,K,J)^*\,
  \Psi^a_i(J)\,\Psi^c_k(J)^*\, \delta_{Ka,c}^{} \,.  $$
The sum over $K$ vanishes unless $a$ and $c$ are on the same
$K$-orbit. But then they are identical, since we work with a definite
set of representatives. Hence there is a contribution from every $K$ that
fixes $a$. In the sum over $J$ all factors $F(a,K,J)$ are equal to 1
by the definition of $\U_a$. Note that the value of $F(a,K,J)$ for
primary fields $a$ that are not fixed points of $K$ is irrelevant. This is
as expected since those values can be modified without affecting
unitarity. The sum over $K$ yields a factor $|\S_a|\delta_{ac}$. Then there
is only a contribution if $a=c$ and hence $\S_a=\S_c$ and
$\U_a=\U_c$. Finally, using orthogonality of the group characters
we get as the final result $\delta_{ac}\delta_{ik}$ as required.

The proofs of the relations $\tilde S^2=\tilde C$ and
$\tilde S\tilde T\tilde S=\tilde T^{-1}\tilde S\tilde T^{-1}$ work in
essentially the same way, except that the cancellation of the factors
$F$ for the intermediate state works
differently here: one now gets a factor $F(b,K,J) \bar F(b,K,J)$
which, however, is again unity because it equals $F(b,K,J)
F(b,K,J^{-1})=1$.

{\small{
Calculation:
  $$\eqalign{\sum_{b,j}&
  \tilde S_{(a,i),(b,j)}T_{(b,j),(b,j)} \tilde S_{(b,j),(c,k)} \cr
  &=\sumRQ b N(a,b)N(b,c)|\U_b| \sum_{J\in \U_b} \Psi^a_i(J) S^J_{ab}
    T_{b,b}^{}\Psi^c_k(J)^* S^J_{b,c}\cr
  &=\sum^{Q=0}_{b} N(a,b)N(b,c)|\U_b|{|\S_b|\over |\G|}
    \sum_{J\in \U_b}\Psi^a_i(J) S^J_{a,b} T_{b,b}\Psi^c_k(J)^*
    S^J_{b,c}\cr
  &=\sum^{Q=0}_{b} {1\over \sqrt{|\U_a||\U_c||\S_a||\S_c|}}\sum_K
    \sum_{J\in \U_b} \Psi^a_i(J) F(a,K,J)^* S^J_{Ka,b}
    T_{b,b}\Psi^c_k(J)^* S^J_{b,c}\cr
  &=\sum_{b} {1\over \sqrt{|\U_a||\U_c||\S_a||\S_c|}} \sum_K
    \sum_{J\in \U_b} \Psi^a_i(J) F(a,K,J)^* S^J_{Ka,b} T_{b,b}
    \Psi^c_k(J)^*  S^J_{b,c}\cr
  &=\sum_{b} {1\over \sqrt{|\U_a||\U_c||\S_a||\S_c|}}
    \sum_K \sum_{J\in \U_a \cap \U_c} \Psi^a_i(J)
    F(a,K,J)^* S^J_{Ka,b} T_{b,b}\Psi^c_k(J)^* S^J_{b,c}\cr
  &={1\over \sqrt{|\U_a||\U_c||\S_a||\S_c|}}\sum_K \sum_{J\in \U_a
    \cap \U_c} \Psi^a_i(J) F(a,K,J)^* T^{-1}_{a,a} S^J_{Ka,c}
    T^{-1}_{c,c}\Psi^c_k(J)^*\cr
  &={1\over \sqrt{|\U_a||\U_c||\S_a||\S_c|}}\sum_K\sum_{J\in \U_a
    \cap \U_c} \Psi^a_{i}(J)T^{-1}_{a,a}\ee^{2\pi\ii Q_K(c)} S^J_{a,c}
    T^{-1}_{c,c}\Psi^c_k(J)^*\cr
  &= {1 \over|\G|}\sum_K T^{-1}_{a,a} \tilde S_{(a,i),(c,k)}
    T^{-1}_{c,c}
   = T^{-1}_{a,a} \tilde S_{(a,i),(c,k)} T^{-1}_{c,c} \,. \cr} $$ }}

For the charge conjugation matrix we find
  $$ \tilde C_{(a,i),(c,k)}=\hat C_{a,c}\,C^c_{i,k} \,, $$
where we defined
  $$ \hat C_{a,c}:={1\over |\S_a|}\sum_{K \in \G} C_{Ka,c} $$
and
  $$ C^c_{i,k} := {1\over |\U_c|} \sum_{J\in \U_c } \Psi^c_i(J)
  \,\eta^J_{c^*}\, \Psi^c_k(J)^* \,. \eqn\Ccik $$
{\small{
Calculation:
  $$\eqalign{\sum_{b,j}&\tilde S_{(a,i),(b,j)} \tilde
    S_{(b,j),(c,k)}\cr
  &=\sumRQ b N(a,b)N(b,c)\,|\U_b| \sum_{J\in \U_b} \Psi^a_i(J)
    S^J_{a,b}
    \Psi^c_k(J)^* S^J_{b,c}\cr
  &=\sum^{Q=0}_{b} N(a,b)N(b,c)\, |\U_b|{|S_b|\over |\G|}
    \sum_{J\in \U_b}\Psi^a_i(J) S^J_{Ka,b} \Psi^c_k(J)^* S^J_{b,c}\cr
  &=\sum^{Q=0}_b {1\over\sqrt{|\U_a||\U_c||\S_a||\S_c|}}\sum_K
    \sum_{J\in \U_b}
    \Psi^a_i(J) F(a,K,J)^* S^J_{Ka,b} \Psi^c_k(J)^* S^J_{b,c}\cr
  &=\sum_{b} {1\over \sqrt{|\U_a||\U_c||\S_a||\S_c|}}
    \sum_K \sum_{J\in \U_b} \Psi^a_i(J)
    F(a,K,J)^* S^J_{Ka,b} \Psi^c_k(J)^* S^J_{b,c}\cr
  &=\sum_{b} {1\over \sqrt{|\U_a||\U_c||\S_a||\S_c|}}
    \sum_K \sum_{J\in \U_a \cap \U_c} \Psi^a_i(J)
    F(a,K,J)^* S^J_{Ka,b} \Psi^c_k(J)^* S^J_{b,c}\cr
  &={1\over \sqrt{|\U_a||\U_c||\S_a||\S_c|}}
    \sum_{K \in \G} \sum_{J\in \U_a \cap \U_c} \Psi^a_i(J)
    F(a,K,J)^* \eta^J_{Ka} C^J_{Ka,c} \Psi^c_k(J)^* \cr
  &={1\over \sqrt{|\U_a||\U_c||\S_a||\S_c|}}
    \sum_{K \in \G} \sum_{J\in \U_a \cap \U_c} \Psi^a_i(J)
    F(a,K,J)^* \eta^J_{a} C^J_{Ka,c} \Psi^c_k(J)^* \cr
  &={1\over |\U_a||\S_a|} \sum_{K \in \G} \sum_{J\in \U_a}
    \Psi^a_i(J)
    F(a,K,J)^* \eta^J_{a} C^J_{Ka,c} \Psi^a_k(J)^* \cr
  &={1\over |\U_a||\S_a|} \sum_{K \in \G} \sum_{J\in \U_a}
    \Psi^a_i(J)
    \eta^J_{a} C^J_{Ka,c} \Psi^a_k(J)^*
   =[{1\over|\S_a|}\sum_{K \in \G}C_{Ka,c}]\times [{1\over |\U_a|}
    \sum_{J\in \U_a} \Psi^a_i(J) \eta^J_{a} \Psi^a_k(J)^*] \cr
  &\equiv [\hat C_{a,c}]\times [{1\over |\U_a|} \sum_{J\in \U_a}
    \Psi^a_i(J) \eta^J_{a} \Psi^a_k(J)^*]
   \equiv \hat C_{a,c}C^c_{i,k} \equiv \tilde C_{(a,i),(c,k)}\,. \cr
  }$$ }}%
The matrix $\tilde C_{(a,i),(c,k)}$ is automatically symmetric since
$\tilde S$ is symmetric. Hence
  $$ \hat C_{a,c}\, C^c_{i,k}=\hat C_{a,c}\, C^a_{k,i} \,. $$
This implies that
  $$  C^c= (C^{c^*})^T \,, $$
a property that can easily be checked explicitly.
The matrix $\tilde C_{(a,i),(c,k)}$ is also automatically unitary,
because $\tilde S$ is. We would like it to satisfy $\unit=\tilde
C^2=\tilde C \tilde C^T$, and hence to be orthogonal. This will be
the case if and
only if it is real, and hence if and only if $C^c$ is real. To verify
this, replace the sum on $J$ by a sum on $J^{-1}$ and use \Inv.
The result is
  $$\eqalign{( C^c_{i,k})^*&=
  {1\over |\U_c|} \sum_{J\in \U_c } \Psi^c_i(J)^*\, (\eta^J_{c^*})^*
  \, \Psi^c_k(J)
  ={1\over |\U_c|} \sum_{J\in \U_c } \Psi^c_i(J^{-1})\,
  \eta^{J^{-1}}_{c^*}\, \Psi^c_k(J^{-1})^*\cr
  & ={1\over |\U_c|} \sum_{J\in \U_c } \Psi^c_i(J)\,
  \eta^{J}_{c^*,c^*} \, \Psi^c_k(J)^*=C^c_{i,k}\,,\cr}$$
so that indeed $\tilde S^4=\tilde C^2=\unit$.

The last condition that $C^c$ should satisfy is that it should be
a permutation. Hence we need
  $${1\over |\U_c|} \sum_{J\in \U_c } \Psi^c_i(J)\, \eta^J_{c^*}
  \,\Psi^c_k(J)^* = \delta_{k,\pi(i)}  $$
for some permutation $\pi$. Multiplying both sides with $\Psi^c_k(K)$
and summing over $k$ yields
  $$ \Psi^c_i(K)\, \eta^K_{c^*} = \Psi^c_{\pi(i)}(K) \eqn\Perm $$
This equation implies that, as a function of $K$, $\eta^K_{c^*}$ is
a ratio of characters of $\U_c$,
which shows that the group property of the $\eta$'s (assumption
$\{5b\}$) is rather natural in this context.
The converse is also true. If $\eta^J$ satisfies $\{5b\}$, then so
does the left hand side of \Perm. Hence the left hand side is itself
a $\U_c$-character, and must be equal to some character $\Psi^c_k$
for some $k$. This is true for all $i$ and defines a permutation
$\pi(i)$.  In other words, the $\eta^J$ are the input which
determines how the  conjugation
acts on the resolved fixed points in the extended theory.

We can also check the fusion rules of additional simple currents
$L$ that are not part of the extension,
      and
are not projected out.
We assume that for such a current condition $\{4\}$ holds, since it
can in principle be used in a further extension\rlap.\foot{If $L$
has fractional spin one may tensor the theory with another theory
that has a complementary simple current, so that the total spin is
integral.}

Now recall that on general grounds we have
$\tilde S_{L(a,i),(b,j)}=\ee^{2\pi\ii Q_L(b)} \tilde S_{(a,i),(b,j)}
$ (see \CSREF). Here the notation $L(a,i)$ stands for
  $$ L(a,i)\equiv(La,Li) \,, $$
with $Li$ an allowed label for the fields into which $La$ is
resolved; thus $L$ acts in the usual way, \ie\ by the fusion product,
on the first label, and as a permutation on the second label.
Substituting this into the formula for $\tilde S$ we obtain
  $$\eqalign{\tilde S_{(La,Li),(b,j)}&=
  {|\G| \over \sqrt{|\U_a|\,|\S_a|\,|\U_b|\,|\S_b|}}
  \sum_{J \in \G} \Psi^a_{Li}(J)\, S^J_{La,b}\, \Psi^b_j(J)^*\cr
  &= {|\G| \over \sqrt{|\U_a|\,|\S_a|\,|\U_b|\,|\S_b|}}
  \sum_{J\in\G} \Psi^a_{Li}(J)\, F(a,L,J)\, \ee^{2\pi\ii
  Q_L(b)}S^J_{a,b}\, \Psi^b_j(J)^*\,.  \cr} $$
Thus we clearly need
  $$ \Psi^a_{Li}(J)\, F(a,L,J)=\Psi_i^a(J)\quad \hbox{~for all~} J
  \,.$$
In other words, for given $F$ there must exist a permutation $L$ of
the labels $i$ such that
this relation holds. By the same arguments as used above for
charge conjugation, this is true if and only if $F$ satisfies
the group property $\{4a\}$ in $J$.
In other words, the numbers $F(\cdot, L, \cdot)$ are the
input which determines how a simple current $L$ acts on the
resolved fixed points in the extended theory.

This shows explicitly that the matrix $\tilde S$ satisfies
conditions [I] -- [VI]. That it also satisfies condition [VII]
essentially follows already from its construction. We will not
give a formal proof here, because condition [VII] (as well as
[VIII]) is of lesser importance. If it were not satisfied,
this would affect the uniqueness of the solution, but we do
not have a proof that our solution is unique anyway.

\section{Fusion rule automorphisms}

\noindent
Next we show that one can identify a symmetry which acts on the
spectrum of the extended theory and is
isomorphic to the discrete abelian group $\G$ of simple
currents by which the chiral algebra was extended. This symmetry
permutes the fields into which a fixed point is resolved. (Thus its
role is analogous to the orbifold group in an (abelian) orbifold
construction, which permutes the various twist sectors of the orbifold.)
This group does not act freely in general, but it acts transitively
     on the fields into which a specific fixed point is resolved
and therefore reflects the `fixed point homogeneity', giving a
concrete meaning to condition [VIII]. The symmetry we consider here
is a special case of the phase rotations of section 5.3; they are now
restricted to act in the same way on all fields, and are required to
satisfy the group property \Dgrp. This ensures that they
       induce
permutations.

Our argument goes as follows. We denote by $\G^*$ the group of
characters of the identification group $\G$, and analogously for
the subgroups $\U_b\subset\G$. Consider now some
fixed point $a$ and take an arbitrary $\G$-character $\phi\in \G^*$.
Then to any character $\Psi^a_i\in\U_a^*$,
$\,\Psi^a_i\!:\,\U_a\to\complex$,
of the untwisted stabilizer $\U_a$ we associate another element
  $$ ^\phi\Psi^a_i:\quad \U_a \to \complex       $$
of $\U_a^*$, which is defined by
  $$ ^\phi\Psi^a_i (J) := \phi(J)\, \Psi^a_i(J) \eqn\pP $$
for all $J\in\U_a$. Combining the group properties of $\phi$ with
respect to $\U_a$ (which are inherited from its group properties
with respect to $\G$)
and the basic properties of the $\U_a$-characters $\Psi^a_i$,
it follows that the functions $^\phi\Psi^a_i$ are again characters of
$\U_a$. As a consequence, \pP\ defines in fact an {\it action}
of $\G^*$ on $\U_a$ and hence permutes the $\U_a$-characters.
(This is completely analogous
to the discussion of charge conjugation and additional simple
currents in the previous section.)
Furthermore, as an abstract group $\G^*$ is isomorphic to $\G$. Since
the primary fields into which a fixed point $a$ is resolved can be
labelled by the elements of the character group $\U^*_a$, we can
therefore describe the situation also as a permutation
  $$  \Psi^a_i \,\mapsto\, ^\phi\Psi^a_i = \Psi^a_{\pi_\phi(i)}
  \eqn\Perm $$
of these fields which is induced by the identification group $\G$.

We claim that when the permutation \Perm\ is performed
simultaneously (with one and the same $\G$-character $\phi$)
on all fixed points, then it
is in fact an automorphism of the fusion rules, i.e.\ we have
  $$ \tilde{\cal N}_{(a,\pi_\phi(i)),(b,\pi_\phi(j))}
  ^{\ \ \ \ \ \ \ \ \ \ \ \ \ \ \ (c,\pi_\phi(k))} =
  \tilde{\cal N}_{(a,i),(b,j)}^{\ \ \ \ \ \ \ \ (c,k)}   \,. \eqn\sym
  $$
Note that the identity is never a fixed point,
\ie\ $\G^*$ acts trivially on it, so that the permutation
$\pi_\phi$ definitely leaves the identity primary field of the
extended theory fixed.

The relation \sym\ can be proven as follows.
We insert the ansatz \XX\ for the $S$-matrix into the Verlinde
formula and perform the resulting summation over $\U_d^*$. This leads
to
  $$ \tilde{\cal N}_{(a,i),(b,j)}^{\ \ \ \ \ \ \ \ (c,k)}
  = \sum_d {1 \over S_{0d}}
  \sum_{{\scriptstyle J\in\U_a\cap\U_d,\atop \scriptstyle
  K\in\U_b\cap\U_d}} S^J_{a,d} S^K_{b,d} {(S^{JK}_{c,d})}^*\,
  \Psi^a_i(J) \Psi^b_j(K) \Psi^c_k(JK)^*_{} \,.\eqn\tN $$
The equality \sym\ then immediately follows with the help of
the group properties of the \G-character $\phi$.

It can happen that the symmetry we just described is not the only
symmetry that acts on the resolved fixed points. As an example,
let us look at the integer spin simple current
invariant of \SU(3) at level 3. It has a single fixed point of order
three. It is easy to check that any permutation of the three
primary fields into which this fixed point is resolved is a
symmetry. Since this modular invariant describes the
conformal embedding of \SU(3) at level 3 into \SO(8) at
level 1, this ${\cal S}_3$-symmetry is nothing else but the
usual triality of \SO(8). More generally, the symmetry is
also enhanced whenever only one non-trivial type of stabilizer
with identical untwisted stabilizer is present. Then any permutation
performed on
all resolved fixed points simultaneously is a symmetry of the theory.

\chapter{WZW-models}

In this section we will introduce the mathematical tools which are
needed to apply our results to WZW-models. In particular, we will
show how to obtain
the matrices $S^J$ in the case of WZW-models. For WZW-models
the action of a simple current corresponds to a certain symmetry of
the Dynkin diagram, and therefore we will have to introduce
some structures which one can associate to any symmetry of a Dynkin
diagram, namely the so-called twining characters and orbit Lie algebras.
For a more detailed explanation, and for some generalizations beyond
the results needed in this paper, we refer to \FSSc\ and \FSSd.

\section{Twining characters}

\noindent
Twining characters are defined in the following way. Consider the
Dynkin diagram of a symmetrizable Kac-Moody algebra. While
the results of \FSSc\ are much more general, for applications in
conformal field theory we can restrict ourselves to untwisted
affine Lie algebras.
An automorphism of such a diagram acts in a canonical way
on the generators corresponding to the simple roots and on the
associated Cartan subalgebra generators. By imposing the automorphism
properties, the action can be extended to all root generators, and
with some additional work also to the derivation, \ie, in physics
terms, the zero-mode Virasoro generator, and even to the whole Virasoro
algebra. In this way we can associate to any
Dynkin diagram automorphism an algebra automorphism \FSSc.
Furthermore the action on the
Cartan subalgebra induces a natural action on the dual space, \ie\
the weight space, and in particular on the highest weights. Integrable
highest weights are either mapped to other integrable
highest weights, or are fixed by the automorphism.

The action of the automorphism $\omega$ on the Kac-Moody algebra and
on the highest weights also induce
an action of $\omega$ on representation spaces, \ie\ $\omega$ can be
implemented by an operator
$\Tau_{\omega}$ which has a well defined action on each state in a
representation space (note that this action is not simply determined
by the weight of the state). This allows to define a new kind of
character, the
{\it twining character}, which is obtained from the usual characters
by inserting $\Tau_{\omega}$ into the trace:
  $$ \chi^{(\omega)}(\tau) = \Tr\, \Tau_{\omega}\, \ee^{2\pi\ii\tau
  L_0} \,,$$
where the trace is over all states in the irreducible representation
space.
Note that $\chi^{(\id)}$ is
       just
the ordinary character. Also, strictly speaking the definition only
makes sense for highest weight representations with a highest weight
that stays fixed under the automorphism, but if the highest weight is not
fixed we can simply set $\chi^{(\omega)} = 0$ for $\omega \not=\id$.

\section{Orbit Lie algebras}

\noindent
Given a diagram automorphism we define a new `folded' Dynkin diagram
which is essentially obtained by identifying nodes
that lie on the same orbit under the automorphism and by
assigning one node of the new diagram to each orbit. The only thing
we then still need is a prescription for the number of lines
between the nodes of the folded diagram. In terms of the Cartan
matrix $\breve A$ of the folded
diagram, this prescription is the following:
  $$ \breve A^{i,j} := s_i \sum_{l=0}^{N_i-1} A^{\omega^l i,j} \,. $$
Here $A$ is the Cartan matrix of the original Lie algebra, and
the labels $i$ and $j$ on the left hand side are (arbitrarily chosen)
orbit representatives. $N_i$ denotes the length of the orbit of node
$i$, and
$\omega$ is the symmetry of the Dynkin diagram.
Finally, the integer $s_i$ is defined as
  $$ s_i:= 1 -\sum_{l=1}^{N_i} A^{\omega^l i,i} \ .  $$

One can show that if $A$ is the Cartan matrix of an affine Lie
algebra, then
so is $\breve A$, except for the order $N$ automorphism of Lie
algebras of untwisted affine type $A_{N-1}$. In
that case there is only a single orbit, and $\breve A$ is formally
the $1\times 1$ matrix $0$. In all other cases
    one has $s_i =1$ or $s_i=2$ for all $i$, and one
    obtains again an affine Lie algebra which
is called the {\em orbit Lie algebra}
      $\breve G$
associated to the
original algebra and its automorphism $\omega$. It can be a twisted
or an untwisted affine Lie algebra.
The main result of \FSSc\ is now that the twining characters are equal to
the ordinary characters of the orbit Lie algebra (this holds for the
full characters, not just for the Virasoro-specialized ones).
 Moreover, in the exceptional case of the order $N$ automorphisms of
 affine $A_{N-1}$, the twining characters can also be computed
 (they are just a power of ${\rm e}^{2\pi\ii\tau}$).

The relation between integrable highest weights of fixed points of
$G$ and integrable highest weights of $\breve G$ is determined
straightforwardly from the folding prescription of the Dynkin diagram.
Dynkin labels $a_i$ of $G$ that are identified by the diagram
automorphism get mapped to a single Dynkin label $a$ on
the corresponding node of the Dynkin diagram of $\breve G$.
This prescription can be extended to include also the Cartan angles
into the twining characters.

\section{Diagram automorphisms corresponding to Simple Currents}

For our purposes we need these results for the special case of
diagram automorphisms related to simple currents.
Simple currents act as a permutation on the highest weight
representations;
the same is true for diagram automorphisms. The correspondence is
not one-to-one: some diagram automorphisms, such as charge
conjugation, do not correspond to simple currents\rlap.\foot{On
the other hand,
there is also an exceptional simple current, namely for
$E_8$ at level 2, for which there is no corresponding diagram
automorphism.}
We will only need simple currents for which there does exist such an
automorphism.
For these currents one finds the following orbit Lie algebras:
\vskip .6truecm
\begintable
algebra $G$ | level | current | spin | orbit Lie algebra $\breve G$ |
level | shift $\delta$ \cr
$A_n^{(1)}$ | ${p(n+1)\over \ell}$ | $J^{\ell}$ |
$\half p (n+1-\ell)$ |  $A_{\ell-1}^{(1)}$ | $p $ |
${[(n+1)^2-\ell^2]\over 24 \ell} p$ \nr
$B_{n+1}^{(1)}$ | $p$ |  $J$ | $\half p$
|  $\tilde B_{n}^{(2)}$ | $p$ | ${2n+3\over 24}p$\nr
$C_{2n}^{(1)}  $ | $p$ | $J$ | $\half pn$ |
 $\tilde B_{n}^{(2)}$ | $p$ | ${5n\over 24} p $ \nr
$C_{2n+1}^{(1)} $ | $2p$ | $J$ | $\half (2n+1)p$ |
$C_n^{(1)}$ | $p$ | ${2n+1\over 8}p$ \nr
$D_n^{(1)}$ | $2p$ | $J_v$ | $p$ |
$C_{n-2}^{(1)}$ | $p$ | ${p\over 4}$ \nr
$D_{2n}^{(1)}$ | $2p$ | $J_s$|$\half pn$
 | $B_n^{(1)}$ | $p$ | ${np\over 8}$ \nr
$D_{2n+1}^{(1)}$ | $4p$ | $J_s$ |$\half(2n+1) p$ | $C_{n-1}^{(1)}$ |
$p$ | ${2n+3\over 8}p$ \nr
$E_6^{(1)}$ | $3p$ | $J$ | $2p$ | $G_2^{(1)}$ | $p$ | $\coeff23 p$\nr
$E_7^{(1)}$ | $2p$ | $J$ | $\coeff32 p$| $F_4^{(1)}$ | $p$ |
$\coeff38 p$\endtable

\noindent
(Here in the first line $\ell$ is a divisor of $n+1$. The orbit Lie
algebra for a power of a current $J$ that has the same order as $J$
is the same as that of $J$.) This table is identical to the one
that has already appeared in \ScYg, except for the second and third
lines. The algebra $\tilde B_n^{(2)}$ is a twisted affine Lie
algebra (we use the notation of \FBk), and appears precisely in those
cases for which no `fixed point theory' could be identified in \ScYg.
The third column specifies the group \G\ of simple currents by means
of that simple current which generates it. For algebras and simple
currents that are not listed in the table, there are no fixed points.
The table in \ScYg\ contained one extra entry for the full $\Zbf_2
\times \Zbf_2$ center of $D_{2n}^{(1)}$. We did not include it here,
because first of all it can be obtained by folding the Dynkin diagram
of one of the orbit Lie algebras once more, and secondly because in
our formalism we
need only fixed point resolution matrices corresponding to cyclic
subgroups of the center, even if the full extension is non-cyclic.
The last column of the table describes the shift $\delta$
in the modular anomaly $h-{c\over 24}$; the entry in the table is
$(h-{c\over 24})|_{G}^{}-(h-{c\over 24})|_{\breve G}$. This shift is
universal in the sense that it only depends on the level of the
algebra $\breve G$, but not on the specific primary field.

Although in the last column we have given the full shift
(except for $\tilde B$, see below), in the following
we need only its fractional part.
     On general grounds (see e.g.\ \ScYg),
this should be
a 12th root of unity for all fixed points of integer spin simple
currents (and hence an odd power of a $24$th root of unity
for fixed points of half-integer spin currents), and indeed it is.

\section{The matrices $S^J$ for WZW-models.}

The orbit Lie algebras appearing in the table all have the property
that their characters (which are equal to the twining characters of
the parent Lie algebra) span a unitary representation of the modular
group (which, incidentally, is no longer true for the other
twisted affine Lie algebras which do not arise in our construction).
This provides us in a natural way with a matrix $S^J$ for every
simple current.

Using the general results of \ScYg\ for fixed point theories, we can
also describe how the shift $\delta$ enters in the
precise definition of the fixed point resolution matrix $S^J$.
Namely, let $\breve  S^J$ be the matrix $S$ of the orbit Lie
algebra corresponding to the folding generated
by $J$; then the matrix $S^J$ that will appear in the fixed point
resolution is given by
  $$ S^J = \ee^{-6\pi\ii\delta} \breve S^J\,. \eqn\ShiftedS $$
The shift $\delta$ ensures that $S^J$ satisfies condition $\{3\}$ of
section 5, \ie\ together with the matrix $T$ of the unextended theory
restricted to the fixed points it forms a representation of the modular group.

Since the twisted algebras $\tilde B_n^{(2)}$ do not correspond
to conformal field theories, the exact definition
of the shift cannot be made directly in terms of differences of
conformal weights, but obviously all we need is that condition
$\{3\}$ is satisfied. The shift $\delta$ is chosen in such a way
that this is the case if we use in \ShiftedS\ for $\breve S^J$ the
Kac-Peterson \KaPe\ formula for the $S$-matrix of
$\tilde B_n^{(2)}$ (or $A_{2n}^{(2)}$ in the
notation of \Kac)\rlap.\foot{One may also compare the formula
for the conformal anomaly given in \Kac\ to the conformal
anomalies of the fixed points. This does not yield
precisely the shift $\delta$.
The difference appears to be due to the fact that
in this case one of the additional arguments (Cartan angles) of
the twining characters is shifted by a constant
with respect to the $A_{2n}^{(2)}$ (orbit Lie algebra) characters as
defined in \Kac. Since modular transformations act not only $\tau$,
but also on the other arguments, this results in an extra
phase in the transformation of the twining characters.}
Since this formula is an essential ingredient in \XX, for
completeness we include it here:
  $$  S_{\Lambda,\Lambda'}= \ii^{|\Delta_+|}
  \left|{M^* \over (k+g) M}\right|^{-\half}
  \sum_{w \in W} \epsilon(w)
  \exp\left[-2\pi\ii\,{(\Lambda+\rho, w( \Lambda'+\rho))\over
  k+g}\right]\ . $$
This formula holds for all algebras in columns 1 or 5 of the table.
To use it one has to select first a definite simple Lie algebra, whose
Dynkin diagram is obtained by removing one node from the Dynkin
diagram of the affine algebra.
For the untwisted Kac-Moody algebras $X^{(1)}$ this is the usual
horizontal subalgebra $X$. For $\tilde B_n^{(2)}$
one chooses the unique $C_n$-type sub-diagram of the Dynkin diagram
to define the horizontal subalgebra.
The level $k$ appearing in the formula is the one appearing in
columns 2 or 6 of the table, and $g$ is the dual Coxeter number,
whose value is $2n+1$ for $\tilde B_n^{(2)}$.
Furthermore $\rho$ is the Weyl vector of the simple
Lie algebra, and $\Lambda$ and $\Lambda'$ are
the restrictions of highest weights of the affine algebra to its
horizontal subalgebra, obtained by ignoring the Dynkin label at
the additional node. For fixed level these restrictions label the
distinct representations uniquely. Finally, $|\Delta^+|$ is the
number of positive roots of the horizontal algebra, the sum is
over the Weyl group of the horizontal algebra weighted
by the determinant $\epsilon(w)$, and the normalization
factor involves the lattice $M$ of the translation subgroup of the
Weyl group of the affine algebra and its dual $M^*$.
For untwisted affine Lie algebras $X^{(1)}$, $M$ is the coroot
lattice of $X$, while for $\tilde B_n^{(2)}$ it is the {\it root}
lattice of $B_n$.

To use the matrices $S^J$ defined by \ShiftedS, one must also
know the precise
relation between the highest weights of the fixed points and
those of the representations of the orbit Lie algebra. This
relation follows in a straightforward way from the folding; it was
already represented diagrammatically in \ScYg, except for the cases
involving $\tilde B_n^{(2)}$, which were interpreted differently
in \ScYg. The prescription for those cases is displayed in the
following diagram:
\vskip .7truecm
\let\picnaturalsize=N
\def\picsize{4.8in}
\def\picfilename{Sres.eps}
\ifx\nopictures Y\else{\ifx\epsfloaded Y\else\input epsf \fi
\let\epsfloaded=Y
\centerline{\ifx\picnaturalsize N\epsfxsize \picsize\fi
\epsfbox{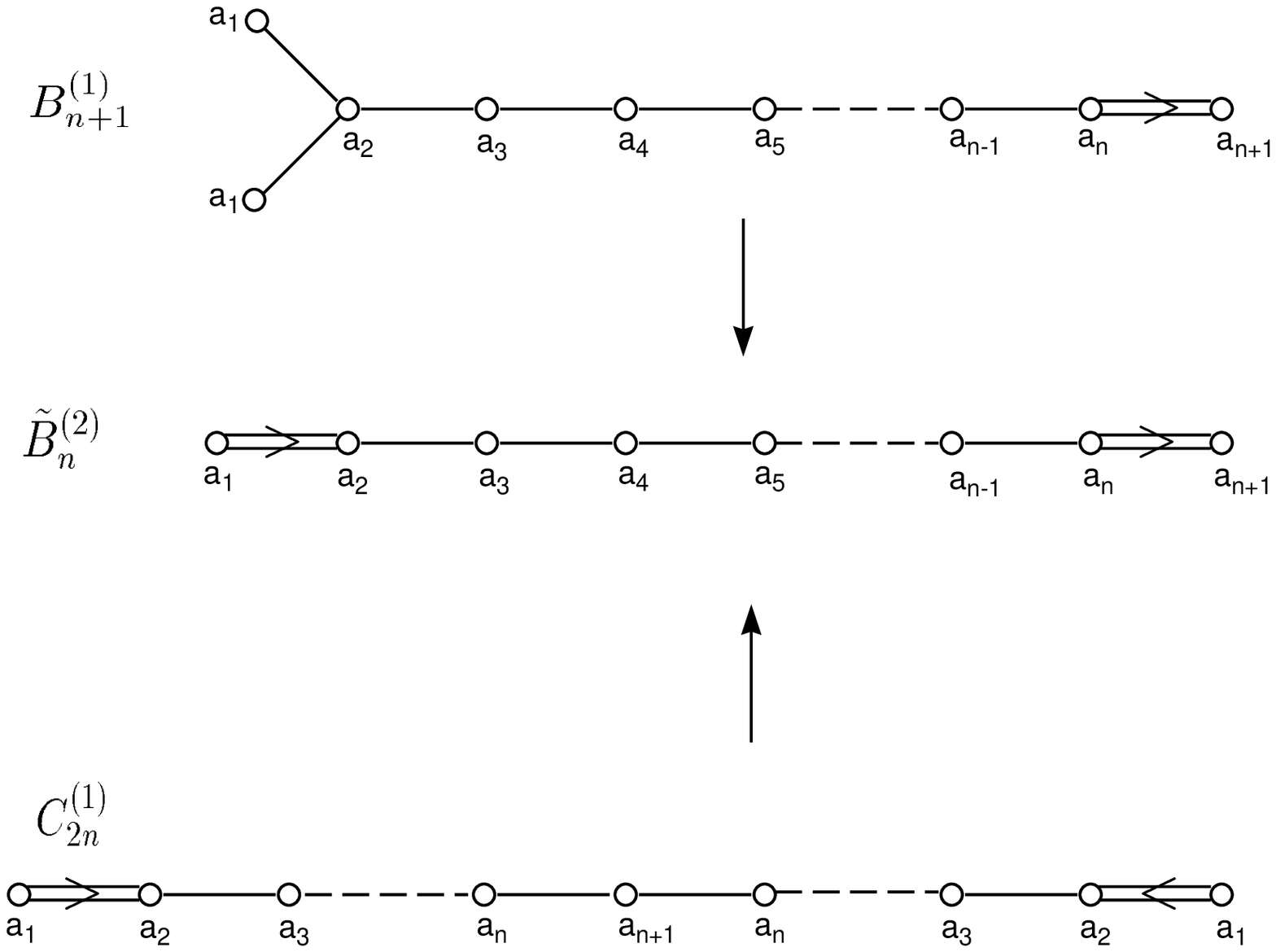}}}\fi
Let us now examine the remaining conditions.
Since we want to do this for arbitrary WZW-models, not just those
based on simple Lie algebras, we will first discuss how the problem
factorizes in the semi-simple case. Consider thus a tensor product
of $N$ simple  WZW-models. We denote their primary fields as $(a_1,
a_2,\ldots, a_N)$. Consider an integer spin simple current $(J_1,
J_2, \ldots,J_N)$. It can only have
fixed points if each of its components $J_i$ does.
Each such component can only have a fixed point if it has integer or
half-integer spin. Note that the conditions $\{1\}$ -- $\{6\}$ were
only formulated for integer spin simple currents $J$ (while the
current $K$ appearing in
    $F(a,K,J)$
can however have any spin).
Hence we will need
to generalize them to half-integer spin currents. This generalization
has to be such that for any integral spin combination the conditions
$\{1\}$ -- $\{6\}$ are automatically satisfied. Note that the need
for such a generalization is not limited to WZW-models, since
the possibility of tensoring exists for any conformal field theory.

It is clear that conditions $\{1\}$ -- $\{6\}$ are satisfied for
                     all extensions of
any tensor product if and only if they are
satisfied for each factor in the tensor product, provided that we
also allow for currents $J$ of half-integer spin.

The matrix $S^J$
acts only on the fixed points of $J$, and therefore condition
$\{1\}$ is satisfied. Furthermore $S^J$ is unitary, so
$\{2\}$ is satisfied as well. Moreover, the matrices $S^J$ are
symmetric, and they are identical
for any two currents $J$ that belong to the same cyclic subgroup
and have the same order, so that $\{6\}$ is satisfied as well.
Condition $\{3\}$ is also satisfied, as pointed out after \ShiftedS\
above.

Unless all fixed points are self-conjugate,
the folded Dynkin diagram
inherits a non-trivial charge conjugation symmetry from the
unfolded one. The matrix $S^J$ satisfies $(S^J)^2=\eta^J C^J$, where
$C^J=\breve C$ is the charge conjugation matrix derived from the folded
diagram. This is thus the unfolded charge conjugation matrix restricted
to the fixed points. Hence the phase $\eta$ is just the square of the
phase appearing in \ShiftedS:
  $$ \eta^J= \ee^{-12\pi\ii\delta(J)} \,. \eqn\DeltaRel $$
For integer spin currents (even those built out of half-integral
spin currents) $\delta(J)$ is a 12th root of unity, and hence
$\eta$ can only be $+1$ or $-1$.
Moreover, it is independent of the fixed points, so that $\{5a\}$ and
$\{5c\}$ are manifestly satisfied. For half-integral spin currents
there is no reason why $\{5c\}$ should be satisfied; indeed in this
case we have $\eta = \pm \ii$, and hence $\{5c\}$ does not hold.

If the fixed points lie on non-trivial orbits of further
simple currents, this will still be the case after folding \FSSb.
Thus the orbit Lie algebra inherits simple current symmetries
from the parent algebra.
By the standard simple current relations for the $S$-matrix (see \CSR),
these simple current symmetries are of the form
  $$ S^J_{Ka,b}=\ee^{2\pi\ii Q^J_K(b)} S^J_{a,b}\,, $$
where
  $$ Q^J_K(b)=h^J(a)+h^J(K^J)-h^J(b) \,. \eqn\QJL $$
Here the superscript $J$ refers to quantities defined directly
in  the orbit Lie algebra. The conformal weights $h^J(a)$ and
$h^J(b)$ differ by a constant (independent of $a$ and $b$)
from the corresponding quantities in the parent algebra; this constant
(\ie\ the shift $\delta$) cancels in \QJL.
   (On the other hand, the primary field $K^J$ is a simple current in
   the orbit Lie algebra. In particular, $K^J$ and $K$ are not related
   the way fixed points are related to fields of the orbit Lie algebra,
   and hence the difference $h^J(K^J)-h(K)$ has nothing to do with the
   shift $\delta$.) The relation
   \QJL\ can thus be rewritten as $Q^J_K(b)=Q_K(b)+h^J(K^J)-h(K)$, so that
     $$  F(a,K,J)=\ee^{2\pi\ii\,[h^J(K^J)-h(K)]} \,. \eqn\FRel$$
Actually the definition
\QJL\ only makes sense if we can interpret the orbit Lie algebra as a
conformal field theory. This is possible in all cases, except for the
$\tilde B$-type algebras, but fortunately the Dynkin diagrams of
those algebras do not possess any symmetries. Once we have a conformal field
theory interpretation, $h^J(K^J)$ is simply the conformal spin
of the simple current corresponding to the symmetry induced by $K$.

We conclude that the phase factor $F(a,K,J)$ is determined by
$h^J(K^J)-h(K)$, and is independent of $a$. Furthermore, since
all matrices are symmetric, we have $F=\bar F$, and then \Fbar\
implies that $F$ is real.
If $K$ is equal to $J$, then $K$ is mapped to the identity
current of the orbit Lie algebra, which maps every field to itself.
In conformal field theories this current is just the identity
primary field, but even for the $\tilde B$-cases we should
interpret $h^J(K^J)$ as zero
for $K=J$ (so that \QJL\ holds with $Q^J_J \equiv 0$ and $a=b$). Hence
    $$ F(J,J) \equiv F(a,J,J)=(-1)^{h(J)} $$
    $$ F(J,J) \equiv F(a,J,J)=(-1)^{h(J)} \eqn\FJJ $$
(see the appendix for a different derivation).

The last requirements to be checked are the group properties of
$\eta$ and $F$. Here we have to pay special attention to currents of
half-integral spin. We note the following.
\item\dash The shifts $\delta$ are an even multiple of $\coeff1{24}$
for integral spin simple
currents and an odd multiple of $\coeff1{24}$ for
half-integral spin currents \ScYg. Therefore $\eta=\pm1$ for integral
spin currents, while $\eta=\pm\ii$ for half-integral spin currents.
\item\dash According to
\Fformh,
  \FJJ,
for half-integral spin currents we
have $F(J,J)=-1$,
while for integral spin currents this quantity is 1.

\noindent
Because of the first point the group property $\{5b\}$ of $\eta$
cannot
be satisfied as it stands for half-integer spin
currents of order 2. The second property implies that integral
spin currents built out of overlapping combinations of
half-integer spin currents (like $K=(J_1,J_2,\one)$ and
$J=(\one,J_2,J_3)$) will have $F(K,J)=-1$.

A check of the properties of $F$ and $\eta$ has to be performed
only at the lowest level  at which a current $J$ has fixed points. This
is because $J$ then has fixed points precisely
at each multiple of that level, and
all relevant quantities (the spins of simple currents and the shifts
$\delta$) are linear in the level. Hence $F$ and $\eta$ scale with powers
of the level. By explicit computation it can be checked
that $\{4a\}$ is satisfied for all WZW-models
with a cyclic center. In that case the phase $F(J,K)$ is
given by \FonOrbit\ for all currents $J$ and $K$ that have
fixed points, and this expression implies $\{4a\}$.

The only simple WZW-theories with a non-cyclic center are those based on
$D_{2n}$. By inspection we see  that $F(J_s,J_v)=F(J_c,J_v)=F(J_s,J_c)=-1$
for $D_{4m}$ at levels $4k+2$, and that $F(J_s,J_v)=F(J_c,J_v)=-1,
F(J_s,J_c)=1$ for $D_{4m+2}$ level $4k+2$ (consequently at levels
that are a multiple of 4 all these quantities are 1). The
values for equal arguments are directly determined, through \FonOrbit,
by the spins of the currents. It is then straighforward to check that
also in this case $\{4a\}$ is indeed satisfied.

As remarked above, the factors $\eta^J$ cannot satisfy $\{5b\}$ for
arbitrary currents $J$. Note however that $\{5b\}$ only needs to hold
on the untwisted stabilizer and only for integer spin currents.
We find that in general instead of  $\{5b\}$ in WZW-models the relation
  $$ \eta^{J_1}\eta^{J_2}=F(J_1,J_2)\,\eta^{J_1J_2} \eqn\Emperical $$
holds. This implies immediately that indeed $\{5b\}$ does hold, because
the factor $F(J,K)$ is equal to 1 for any two currents in the
untwisted stabilizer by the definition of the latter. This is in fact
a non-trivial check on the applicability of our formalism to
WZW-models. Although establishing \Emperical\ empirically is
sufficient for our purpose, a deeper insight would be welcome. Note
that {\it a priori} the quantities $F$ and $\eta$ are not in any obvious
way related. The relation \Emperical\ as well as the fact that $F(a,K,J)$ is
independent of $a$ suggest that $F$ plays a more fundamental role than we
described so far. We comment briefly on this is in the conclusions.

Finally we give two examples where the untwisted stabilizer
is smaller than the stabilizer. First consider
the following modular invariant of $D_4$ level 2:
  $$ |\X_\one + \X_v+\X_s+\X_c |^2 + 4\, |\X_{28}|^2 \,. $$
Here the subscripts denote, respectively, the identity field, the
three simple currents $J_v$, $J_s$ and $J_c$, and the unique
primary field with ground state dimension 28 that is fixed
by all currents. Because the currents have spin 1, this is a
conformal embedding, and it can be interpreted in terms of
$E_7$ level 1, provided we take the factor 4 inside the square.
This provides the correct ground state dimension 56 for the
$E_7$-representation, and it yields the correct total number of
primaries, namely 2. The stabilizer of the $(28)$ is $\S_{28}
=\{\one ,J_v,J_s,J_c\}$, but the untwisted stabilizer is just the
identity, $\U_{28}=\{\one\}$.

An example with reduction of the stabilizer due to spin-1/2 currents
is also easy to find. Consider the product of three $B_n$ WZW-theories
at level 1. We may extend the algebra by the currents
$(J,J,\one)$, $(J,\one,J)$ and $(\one,J,J)$. This extension has
again an interpretation as a conformal embedding, namely as
$\SO(d_1)\oplus \SO(d_2)\oplus \SO(d_3) \subset \SO(d_1+d_2+d_3)$, with
all $d_i$ odd. The spinor of each $\SO(d_i)$ is fixed by the simple
current $J$, and hence the product of the three spinors has
a stabilizer of order 4. But $\SO(d_1+d_2+d_3)$ has only one spinor
representation, so that once again we see that the factor 4
must be absorbed into the character. In this way one gets
the correct normalization
      $2^{(d_1+d_2+d_3 -1)/2}$
of the ground state.

We have applied our formula to many other cases, in particular
with the aim of checking the fusion rules. In all cases the
fusion coefficients were positive integers, but a proof of this property
is still lacking. A computer program that implements our formalism and
calculates the (conjectured) fusion coefficients and the
 matrix $\tilde S$
for any
simple current extended WZW-model is available from the second author.

\chapter{Conclusions}

In this paper we have presented strong evidence for the formula \XX\
which describes the modular $S$-matrix for extensions of the chiral
algebra by integer spin simple currents. The basic ingredients leading
to this formula are the following. For any simple current $J$ in the
extension we postulate the existence of a representation of the
modular group defined on
the fixed points only, leading to the matrices $S^J$. These
matrices determine a subgroup of the stabilizer of each fixed point,
the so-called untwisted stabilizer. A formalism could then be
developed which  expresses
the matrix $\tilde S$ in terms of the matrices $S^J$ and the
group characters of the untwisted stabilizer groups $\U_a$.

Some important open problems remain. We have not shown that our
solution for $\tilde S$ always leads to non-negative integral
fusion coefficients, not to mention even the existence of a
conformal field theory for each
extension. We have also not shown that our solution is unique. The
arguments in section 4 suggest strongly that it is the only solution
to the conditions [I] -- [VIII], given the matrices $S^J$, but
uniqueness of the
latter is by no means proved. Nevertheless, the existence of natural
candidates (the modular matrices for torus one-point functions in
general, and $S$-matrices of orbit Lie algebras for WZW-models)
suggests that our solution is unique in this respect.

In some situations fixed point resolutions
which are still more general than the ones considered here
can work, which however violate the postulate of fixed point
homogeneity. In
this case equation \smmm\ is replaced by a more general decomposition
of $|\S_a|$ into squares. An example we have checked is the
simple current invariant of $A_4$ level 5. If one splits the fixed
point using $m_1=2,\, m_2=1$ instead of the homogeneous splitting $m_i=1,
\,i=1,2,...\,,5$, and uses the same matrix $S^J$, then one arrives
at a matrix $\tilde S$ that satisfies all conditions (except
homogeneity).
Furthermore both resolutions yield correct fusion rules
 for a different number of primary fields.
Of course this does not mean that both yield a conformal field theory.

It is tempting to speculate and interpret our results directly
in terms of the underlying
extension of the chiral algebra $\cal A$. The simple currents used in
the extension form a group $\G$. One would like to construct the
extended chiral algebra
      $\tilde{\cal A}$
in terms of this group $\G$ and the original
chiral algebra $\cal A$. More precisely, instead of the group $\G$ one
should work with its group algebra ${\cal A}_\G$. The fact that we
have to distinguish between the ordinary stabilizers and the untwisted
stabilizers shows, however, that this idea is somewhat
too naive. Interestingly enough, there is a generalization of the
notion of a group algebra, the so-called twisted group algebras.
The twisted group algebra $\Gt$ associated to $\G$ is still
associative, as required for a chiral algebra, but no longer
commutative, even though $\G$ is.
The representation theory of $\Gt$ precisely fits with what we expect
for
      $\tilde{\cal A}$
from our results. In particular,
the irreducible representations of $\Gt$ are not one-dimensional
any more; they are labelled by the characters of a subgroup $\U$ of
$\G$ and have dimension $\sqrt{|\G|/|\U|}$; this should explain the
prescription \smmm\ for decomposing the stabilizer groups.

Unfortunately, such an explicit description of
      $\tilde{\cal A}$
is still
lacking. A first step in this direction would be an explicit
construction of the chiral vertex operators corresponding to simple
currents. In the case of WZW-theories,
this should in particular lead to a proof of our expectation that the
matrices $S^J$ describe the modular transformation properties of
the one-point  functions of simple currents on the torus.
Such a construction would also justify our assumption on the
existence of the matrices $S^J$ for arbitrary conformal field theories.
Note that in the case of WZW-theories one can identify the matrices
$S^J$ with the $S$-matrices of the relevant orbit Lie algebras,
and therefore all ingredients for the $S$-matrix formula are known.

Let us also mention possible applications of our formula beyond
conformal field theory. When combined with Verlinde's formula,
the Kac-Peterson formula for the $S$-matrix
allows to compute the rank of certain vector bundles over the moduli
space of semi-stable bundles with structure group $G$, where $G$ is
a simply connected compact Lie group. Since integer spin simple
current invariants can be interpreted in terms of WZW-theories on
non-simply connected compact Lie groups \doubref\AbGe{\fegk}, it is
reasonable to expect that our formula, combined again with the
Verlinde formula,
can be used to compute analogous quantities for the case when the
structure group $G$ is not simply connected.

\ack

One of us (A.N.S) was 
``Profesor Visitante IBERDROLA de Ciencia y Tecnologia"
at the ``Universidad Autonoma",
Madrid while most of this work was done. He would
like to thank the theory group of Autonoma University for their
hospitality and Iberdrola for making this visit possible.

\vfill\eject
\appendix
\leftline{{\sl Properties of $F$ and $\eta$}}

There are several restrictions on $F$ and $\eta$ due to conditions
$\{1\}$ -- $\{6\}$\rlap,\foot{Curly brackets refer to the conditions
stated in
        section
5.}
and in addition there is a residual transformation that
respects all conditions. Here we will use the latter to restrict
the values of $F$ and $\eta$.

First of all, because of the unitarity condition $\{2\}$ we have
  $$ \eqalign{1&=\sum_b S^J_{Ka,b}(S^J_{b,Ka})^\dagger\cr
  &= \sum_b F(a,K,J) F(a,K,J)^* S^J_{a,b}(S^J_{a,b})^*\
  =  F(a,K,J) F(a,K,J)^*,\cr} $$
so that $F$ is a phase. Note that for this conclusion it is essential
that the numbers $F$ do not depend on $b$.
Furthermore we have manifestly $F(a,\one,J)=F(a,K,\one)=1$.

Consider now first the values of $F(a,K,J)$ when $a$ is a fixed
point of $K$. Then condition $\{4\}$ implies that either $F(a,K,J)
=\ee^{-2\pi \ii Q_K(b)}$ or $S^J_{a,b}=0$. Since $S^J_{a,b}$ cannot
vanish
for all $b$ because of unitarity, this implies that
  $$ F(a,K,J) = \ee^{2\pi \ii\, q(a,K,J)}\,, $$
where $q(a,K,J)$ is an allowed charge with respect to the current
$K$, \ie\
an $M$th root of unity, where $M$ is the order of $K$ in $\G$.

If $K=J$ (or a power of $J$) we have $Q_J(b)=0=Q_K(b)$, because $b$
is a fixed point of the (integer spin) simple current $J$.
It follows that $F(a,J^n,J)=1$. Hence
by the group property in the third argument we have $F(a,J^n,J^m)=1$
for all $n$ and $m$. Thus $F$ is trivial within cyclic groups generated
by integer spin currents.

The matrices $S^J$ can also be defined for half-integer spin
currents, the only currents of fractional spin that can possess
fixed points. In that
case the fixed points of $J$ have charge $\half$ with respect to $J$,
and the foregoing result generalizes to
  $$ F(a,J^n,J^m)=(-1)^{nm\,h(J)} \ , \eqn\FonOrbit $$
where $h(J)$ is the spin of $J$. Any half-integer spin simple current
is local with respect to itself. Therefore $h(J^m)=mh(J) \mod 1$.
Relation \FonOrbit\ thus implies that for currents $J$ and $K$ from
the same cyclic orbit, $F(a,J,K)$  is equal to $-1$ if both $J$ and $K$
have half-integer spin, and equal to 1 otherwise.

Now we compute the value of $F(a,K,J)$ on the $\G$-orbit
through $a$, still for the case $Ka=a$.
Suppose that $L$ is a current in $\G$. Then
  $$S^J_{KLa,b}=F(La,K,J)\, \ee^{2\pi \ii Q_K(b)} S^J_{La,b}=
  F(La,K,J)F(a,L,J)\, \ee^{2\pi \ii [Q_K(b)+Q_L(b)]} S^J_{a,b} \,.
\eqn\Forbit$$
Using commutativity of the currents we have also
  $$S^J_{LKa,b}=F(Ka,L,J)\,\ee^{2\pi \ii Q_L(b)} S^J_{Ka,b}=
  F(Ka,L,J)F(a,K,J)\,\ee^{2\pi \ii [Q_K(b)+Q_L(b)]} S^J_{a,b} \,.
\eqn\Forbita$$
If $K$ fixes $a$, then when equating \Forbit\ and \Forbita\
the factor $F(a,L,J)$ cancels out, and
(choosing $b$ such that $S^J_{a,b}\not = 0$, which is again possible
because $S^J$ is unitary) we find
  $$F(La,K,J)=F(a,K,J)\ ,\eqn\OrbitConst $$
so that $F(a,K,J)$ is constant on all simple current orbits of $a$.
A third way of writing $S^J_{KLa,b}$ is as follows:
  $$S^J_{LKa,b}= F(a,LK,J)\,\ee^{2\pi\ii[Q_K(b)+Q_L(b)]} S^J_{a,b}
  \,;$$
comparing with \Forbit\ or \Forbita\ we see that
  $$ F(a,KL,J)=F(a,K,J)F(a,L,J) \eqn\FirstGrp $$
if $a$ is a fixed point of $K$ as well as $L$.

      Next we recall from section 5 that
conditions $\{1\}$ -- $\{6\}$ remain valid if we
transform $S^J$ as in \PhaseRot, provided that the numbers $D^J$
satisfy \DJinv\ and condition \Dgrp, or at least a sufficient subset
of the latter to
preserve the group properties of $\eta^J$ and $F(a,K,J)$.
This transformation changes both $F$ and $\eta$, except the
value of $F(a,K,J)$ if $Ka=a$ and the value of $\eta$ on
self-conjugate fields. The phase factor $F(a,K,J)$ changes to
$D^J_{a}(D^J_{Ka})^*F(a,K,J)$, where
$D_a$ denotes the diagonal matrix element $D_{a,a}$. We will use
the analogous notation also for $\eta$. On the $K$-orbit of $a$ we
then have
  $$F'(K^{\ell}a,K,J)=D^J_{K^{\ell}a}(D^J_{K^{\ell+1}a})^*
  F(K^{\ell}a,K,J) \,. \eqn\Eqns$$
The matrices $D^J$ can be used to make all $F'$ equal to a constant
$X$ on the
$K$-orbit of $K$ (this works for all currents $K$ simultaneously, as
can \eg\ be checked by considering one generator for each cyclic
subgroup). This can be seen as follows. Suppose $N$ is the smallest
positive integer for which $K^N a = a$.
Multiplying the equations for $\ell=0,1,\dots, N-1$ we then find
  $$ X^N = \prod_{\ell=0}^{N-1} F(K^{\ell}a,K,J) \,. \eqn\Cdef $$
The right hand side can be computed as
  $$ \eqalign{S^J_{K^Na,b}&=F(K^{N-1}a,K,J)\,\ee^{2\pi \ii Q_K(b)}
    S^J_{K^{N-1}a,b}\cr
  &=F(K^{N-1}a,K,J)\,\ee^{2\pi \ii Q_K(b)}F(K^{N-2}a,K,J)\,\ee^{2\pi
  \ii Q_K(b)}    S^J_{K^{N-2}a,b}\cr &=\,\ldots\,
   =\ee^{2\pi\ii N Q_K(b)}\prod_{\ell=0}^{N-1}
  F(K^{\ell}a,K,J)S^J_{a,b} \,.\cr} \eqn\CDefa $$
On the other hand we have
  $$ S^J_{K^Na,b}=\ee^{2\pi \ii N Q_K(b)}F(a,K^N,J)\,S^J_{a,b}\,.
  \eqn\CDefb $$
Comparing \Cdef, \CDefa\ and \CDefb\ we see that $X$ is
an $N$th root of $F(a,K^N,J)$. The latter quantity has
been discussed already above, since $K^N$ fixes $a$.
For any valid choice of $X$ the equations \Eqns\ then
allow us to express the coefficients $D^J_{Ka}$ recursively in
terms of $D^J_{a}$. Since only the ratios enter, the latter remains
a free parameter.
It is natural to choose the values of $X$ in such a way that
the group property \FirstGrp\ of the $F$ coefficients is respected
for all primaries $a$, not just for fixed points of $K$. This can be
achieved (although not uniquely) by choosing the phase $q(a,K,J)$ in
the range $0 \leq q < 1$ and defining
$$ F(a,K^{\ell},J)=\exp[{2\pi\ii{\ell\over N}\, q(a,K^N,J)}] \,. $$
Note that if $K^M=\one$, then $(K^N)^{M/N}=\one$, and then $q$ is
an integer multiple of $N\over M$. This implies that
$F(a,K^M,J)=1$, as is clearly required.
This definition can be directly generalized to arbitrary currents
$K\in \G$, namely as
  $$ F(a,\prod_i K_i^{\ell_i},J)=
  \exp[2\pi\ii\sum_i{\ell_i\over N_i}\, q(a,K_i^{N_i},J)] \,,
  \eqn\Fdef $$
which for given $J$ manifestly has the required group property
in the second argument. On the other hand, the group property
$\{4a\}$ in the third argument is not necessarily respected for all
$J$. This is because in solving the equations \Eqns\ we did not
pay attention yet to \Dgrp, which does not allow us an
arbitrary choice of the $N$th root of unity for each $J$.
This problem can be solved by choosing \Fdef\ on
a set of generators $J_i$ of $\S_a$, and extending it over $\S_a$
with the help of the group property.

Since $F(a,K,J)=1$ if $K$ and $J$ belong to the same cyclic
group, let us assume now that they are generators of two different
cyclic groups $\Zbf_N$ and $\Zbf_M$, respectively. Then
$F(a,K,J)=\Phi_a$, where $\Phi_a$ is a phase factor, with
$(\Phi_a)^{\GCD(N,M)}=1$ ($\GCD(N,M)$ denotes the greatest common
divisor of $N$ and $M$).
       If $J$ and $K$ have integral spin, then
using the group property in the last argument and
$F(a,K,K^n)=1$ we find
that $F(a,K,K^nJ^m)=(\Phi_a)^m$. Now
we may use the group property in the second argument as well as
$F(a,(K^nJ^m)^{\ell},K^nJ^m)=1$ to derive
$F(a,K^t(K^nJ^m)^{\ell},K^nJ^m)=(\Phi_a)^{tm}$.
The power of $\Phi_a$ can be written as $tm=pm-nq$ with
$p:=t+n\ell$ and $q:=m\ell$. The final result is thus
  $$ F(a,K^pJ^q,K^nJ^m)=(\Phi_a)^{pm-nq} \,. \eqn\Fform $$
This implies in particular the relation
  $$ F(a,J_1,J_2)=F(a,J_2,J_1)^*\eqn\Fsym $$
for arbitrary currents $J_1$ and $J_2$.

For mutually local\foot{Note that mutually non-local currents
cannot have simultaneous
fixed points, since the action of one current changes the charge
with respect to the other. The definition of $F$ is only
relevant for two currents that fix the same field.}
currents $K$ and $J$ of arbitrary integer or half-integer spin
the foregoing result generalizes to
  $$ F(a,K^pJ^q,K^nJ^m)=(\Phi_a)^{pm-nq}(-1)^{pnh(K)+qmh(J)}\,, \eqn\Fformh $$
which again implies $\Fsym$.

This concludes the discussion of $F$. Having fixed $F$, we also know
$\bar F$ using $\{4\}$ and $\{6\}$:
  $$ \bar F(a,K,J)=F(a,K,J^{-1})=F(a,K,J)^*  \,.\eqn\Fbar $$

Now consider the phases $\eta$.
On conjugate representations $a$ and $c=a^*$ we have
$(Ka)^*=K^{-1}c$, so that
  $$\eqalign{ C^J_{a,c}&=C^J_{Ka,K^{-1}c}=(\eta^J_{Ka,Ka})^* \sum_b
  S^J_{Ka,b}S^J_{b,K^{-1}c}\cr &=F(a,K,J)\bar
  F(c,K^{-1},J)(\eta^J_{Ka})^*\eta^J_{a}\,C^J_{a,c}\,.}\eqn\CCC $$
If $a$ is a fixed point of $K$, then the factors $\eta^J$ cancel, and
we find
  $$ F(a,K,J)=\bar F(a^*,K^{-1},J)^*=F(a^*,K^{-1},J)=F(a^*,K,J)^* \,.
  \eqn\FconjOne $$
Because of the definition \Fdef\ of $F$ for any $a$
and the way this definition was extended over $\S_a$, this
relation holds automatically for all values of $a$ if it holds for
fixed points of $K$. Then the factors $F$ in \CCC\ cancel for all
$a$, and hence the factors $\eta$ must cancel as well. Thus we get
  $$ \eta^J_{Ka}=\eta^J_a  \hbox{~~~for all~ $K \in \G$} \,. $$

Under the transformation \PhaseRot\ $\eta_{a}$ is changed to
$\eta_{a}D^J_{a}(D^J_{a^*})^*$, and $\eta_{a^*}$ is
transformed by the complex conjugate of this factor. If $a$ and $a^*$
lie on different orbits, this allows us to transform both
$\eta_a$ and $\eta_a^*$ to the value 1. Since we have already made
the value of $\eta$ constant on $\G$-orbits, we thus need just one
parameter per conjugate pair to fix $\eta_a$. This parameter
is precisely available, because one parameter $D^J_a$ per orbit
remained free. On the other hand, after fixing $\eta$, there is still
one free parameter per pair of conjugate orbits. The group property
$\{5b\}$ for
$\eta$ is manifestly satisfied if we set $\eta^J$ to 1 for all $J$.

If $a$ and $a^*$ lie on the same orbit, we have
$\eta_a=\eta_{a^*}=(\eta_a)^*$ so that $\eta_a = \pm1$. We do not
have
any free parameters to change this value, since the ratio
$D_a/D_{a^*}$ is already fixed. Note that even if an orbit is
self-conjugate,
it may nevertheless happen that it does not contain any
self-conjugate field\rlap.\foot{Consider for example a current of
order 2 in $\SU(4)$ level 4. The fields (3,0,1) and (1,0,3) are
conjugate and lie on the same orbit.} Therefore this conclusion does
not follow directly from $\{5c\}$. Note that the group property for
$\eta$, $\{5b\}$, is
satisfied because the group property of $D^J$ was respected.

\par \penalty-4000\vskip\chapterskip
   \spacecheck\referenceminspace \immediate\closeout\referencewrite
   \referenceopenfalse
   \line{\fourteenrm\hfil REFERENCES\hfil}\vskip\headskip
   \endlinechar=-1
   \input referenc.texauxil
   \endlinechar=13
   
\end